\documentclass[aps,prl,superscriptaddress,floatfix]{revtex4}
\usepackage{amsfonts}
\usepackage{epsfig}
\usepackage{amsmath,amssymb}
\usepackage{times}
\usepackage{xcolor}

\begin{document}
\title{Asymmetrically interacting spreading dynamics on complex layered networks}

\author{Wei Wang}
\affiliation{Web Sciences Center, University of Electronic
Science and Technology of China, Chengdu 610054, China}

\author{Ming Tang}\email{tangminghuang521@hotmail.com}
\affiliation{Web Sciences Center, University of Electronic
Science and Technology of China, Chengdu 610054, China}
\affiliation{Center for Atmospheric Remote Sensing(CARE),
Kyungpook National University, Daegu, 702-701, South Korea}

\author{Hui Yang}
\affiliation{Web Sciences Center, University of Electronic
Science and Technology of China, Chengdu 610054, China}

\author{Younghae Do}
\affiliation{Department of Mathematics, Kyungpook National
University, Daegu 702-701, South Korea}

\author{Ying-Cheng Lai}
\affiliation{School of Electrical, Computer and Energy Engineering,
Arizona State University, Tempe, Arizona 85287, USA}

\author{GyuWon Lee}
\affiliation{Department of Astronomy and Atmospheric Sciences,
Center for Atmospheric Remote Sensing(CARE), Kyungpook National
University, Daegu, 702-701, South Korea }

\begin{abstract}
The spread of disease through a physical-contact network and the spread
of information about the disease on a communication network are two
intimately related dynamical processes. We investigate the asymmetrical
interplay between the two types of spreading dynamics, each occurring
on its own layer, by focusing on the two fundamental quantities underlying
any spreading process: epidemic threshold and the final infection ratio.
We find that an epidemic outbreak on the contact layer can induce an
outbreak on the communication layer, and information spreading can
effectively raise the epidemic threshold. When structural correlation
exists between the two layers, the information threshold remains unchanged
but the epidemic threshold can be enhanced, making the contact layer more
resilient to epidemic outbreak. We develop a physical theory to understand
the intricate interplay between the two types of spreading dynamics.
\end{abstract}

\date{\today}
\maketitle

Epidemic spreading~\cite{Pastor-Satorras:2001,Tang:2009,Shu:2012,
Wang:2013,Gross:2006,Holme:2012} and information
diffusion~\cite{Zanette:2002,Liu:2003,Zhou:2007,Noh:2004}
are two fundamental types of dynamical processes on complex networks.
While traditionally these processes have been studied independently,
in real-world situations there is always coupling or interaction
between them. For example, whether large-scale outbreak of a disease
can actually occur depends on the spread of information about the disease.
In particular, when the disease begins to spread initially, individuals
can become aware of the occurrence of the disease in their neighborhoods
and consequently take preventive measures to protect themselves. As a
result, the extent of the disease spreading can be significantly
reduced~\cite{Funk:2010-2,Meloni:2011,Wang:2012}. A recent example is
the wide spread of severe acute respiratory syndrome (SARS) in China in
2003, where many people took simple but effective preventive measures
(e.g., by wearing face masks or staying at home) after becoming aware of
the disease, even before it has reached their neighborhoods~\cite{Tai:2007}.
To understand how information spreading can mitigate epidemic outbreaks,
and more broadly, the interplay between the two types of spreading
dynamics has led to a new direction of research in complex network
science~\cite{Manfredi:2013}.

A pioneering step in this direction was taken by Funk \emph{et al.}, who
presented an epidemiological model that takes into account the spread of
awareness about the disease~\cite{Funk:2009,Funk:2010-1}. Due to information
diffusion, in a well-mixed population, the size of the epidemic outbreak
can be reduced markedly. However, the epidemic threshold can be enhanced
only when the awareness is sufficiently strong so as to modify the key
parameters associated with the spreading dynamics such as the infection
and recovery rates. A reasonable setting to investigate the complicated
interplay between epidemic spreading and information diffusion is to assume
two interacting network layers of of identical set of nodes, one for each
type of spreading dynamics. Due to the difference in the epidemic and
information spreading processes, the connection patterns in the two layers
can in general be quite distinct. For the special case where the two-layer
overlay networks are highly correlated in the sense that they have completely
overlapping links and high clustering coefficient, a locally spreading
awareness triggered by the disease spreading can raise the threshold even
when the parameters in the epidemic spreading dynamics remain
unchanged~\cite{Funk:2009,Funk:2010-1}. The situation where the two
processes spread successively on overlay networks was studied
with the finding that the outbreak of information diffusion can constrain
the epidemic spreading process~\cite{Funk:2010-3}. An analytical approach
was developed to provide insights into the symmetric interplay between
the two types of spreading dynamics on layered networks~\cite{Marceau:2011}.
A model of competing epidemic spreading over completely overlapping networks
was also proposed and investigated, revealing a coexistence regime in which
both types of spreading can infect a substantial fraction of the
network~\cite{Karrer:2011}.

While the effect of information diffusion (or awareness) on epidemic spreading
has attracted much recent interest~\cite{Ahn:2006,Kiss:2010,Perra:2011,
Sahneh:2012,Wu:2012,Ruan:2012,Jo:2006,Wang:2012_2}, many outstanding issues
remain. In this paper we address the following three issues. The first
concerns the network structures that support the two types of spreading
dynamics, which were assumed to be identical in some existing works.
However, in reality, the two networks can differ significantly in their
structures. For example, in a modern society, information is often
transmitted through electronic communication networks such as
telephones~\cite{Jiang:2013} and the Internet~\cite{Faloutsos:1999},
but disease spreading usually takes place on a physical contact
network~\cite{Starnini:2013}. The whole complex system should then be
modeled as a double-layer coupled network (overlay network or
multiplex network)~\cite{Buldyrev:2010,Gao:2012,Cozzo:2013,Kivela:2013,Kim:2013},
where each layer has a distinct internal structure and the interplay between
between the two layers has diverse characteristics, such as
inter-similarity~\cite{Parshani:2010}, multiple support
dependence~\cite{Shao:2011}, and inter degree-degree
correlation~\cite{Lee:2012}, etc. The second issue is that
the effects of one type of spreading dynamics on another are typically
asymmetric~\cite{Ahn:2006}, requiring a modification of the symmetric
assumption used in a recent work~\cite{Marceau:2011}. For example, the
spread of a disease can result in elevated crisis awareness and thus
facilitate the spread of the information about the disease~\cite{Funk:2010-1},
but the spread of the information promotes more people to take preventive
measures and consequently suppresses the epidemic spreading~\cite{Ruan:2012}.
The third issue concerns the timing of the two types of spreading dynamics
because they usually occur simultaneously on their respective layers and
affect each other dynamically during the same time period~\cite{Marceau:2011}.

Existing works treating the above three issues separately showed that each
can have some significant effect on the epidemic and information spreading
dynamics~\cite{Funk:2009,Marceau:2011,Mills:2013}. However, a unified
framework encompassing the sophisticated consequences of all three issues
is lacking. The purpose of this paper is to develop an asymmetrically
interacting spreading-dynamics model to integrate the three issues so as
to gain deep understanding into the intricate interplay between the epidemic
and information spreading dynamics. When all three issues are taken into
account simultaneously, we find that an epidemic outbreak on the contact
layer can induce an outbreak on the communication layer, and information
spreading can effectively raise the epidemic threshold, making the contact
layer more resistant to disease spreading. When inter-layer correlation
exists, the information threshold remains unchanged but the epidemic
threshold can be enhanced, making the contact layer more resilient to
epidemic outbreak. These results are established through analytic theory
with extensive numerical support.

\section*{Results}

In order to present our main results, we describe our two-layer network
model and the dynamical process on each layer. We first treat the case
where the double-layer networks are uncorrelated. We then incorporate
layer-to-layer correlation in our analysis.

\textbf{\emph{Model of communication-contact double-layer network}.}
Communication-contact coupled layered networks are one class of multiplex
networks~\cite{Gomez:2013}. In such a network, an individual (a node)
not only connects with his/her friends on a physical contact layer
(subnetwork), but also communicates with them through the (electronic)
communication layer. The structures of the two layers can in general
be quite different. For example, an indoor-type of individual has few
friends in the real world but may have many friends in the cyber space,
leading to a much higher degree in the communication layer than in
the physical-contact layer. Generally, the degree-to-degree correlation
between the two layers cannot be assumed to be strong.

Our correlated network model of communication-contact layers is
constructed, as follows. Two subnetworks $A$ and $B$ with the same node
set are first generated independently, where $A$ and $B$ denote the
communication and contact layers, respectively. Each layer possesses
a distinct internal structure, as characterized by measures such as
the mean degree and degree distribution. Then each node of layer $A$
is matched one-to-one with that of layer $B$ according to certain rules.

In an uncorrelated double-layer network, the degree distribution of one layer
is completely independent of the distributions of other layer. For
example, a hub node with a large number of neighbors in one layer is not
necessarily a hub node in the other layer. In contrast, in a correlated
double-layer network, the degree distributions of the two layers are strongly
dependent upon each other. In a perfectly correlated double-layer network,
hub nodes in one layer must simultaneously be hub nodes in the other layer.
Quantitatively, the Spearman rank correlation coefficient~\cite{Cho:2010,Lee:2012}
$m_s$, where $m_s\in[-1,1]$ (see definition in {\bf Methods}), can be
used to characterize the degree correlation between the two layers.
For $m_s>0$, the greater the correlation coefficient, the larger
degree a pair of counterpart nodes can have. For $m_s < 0$, as $|m_s|$
is decreased, a node of larger degree in one layer is matched with a
node of smaller degree in the other layer.

\begin{figure}
\begin{center}
\epsfig{file=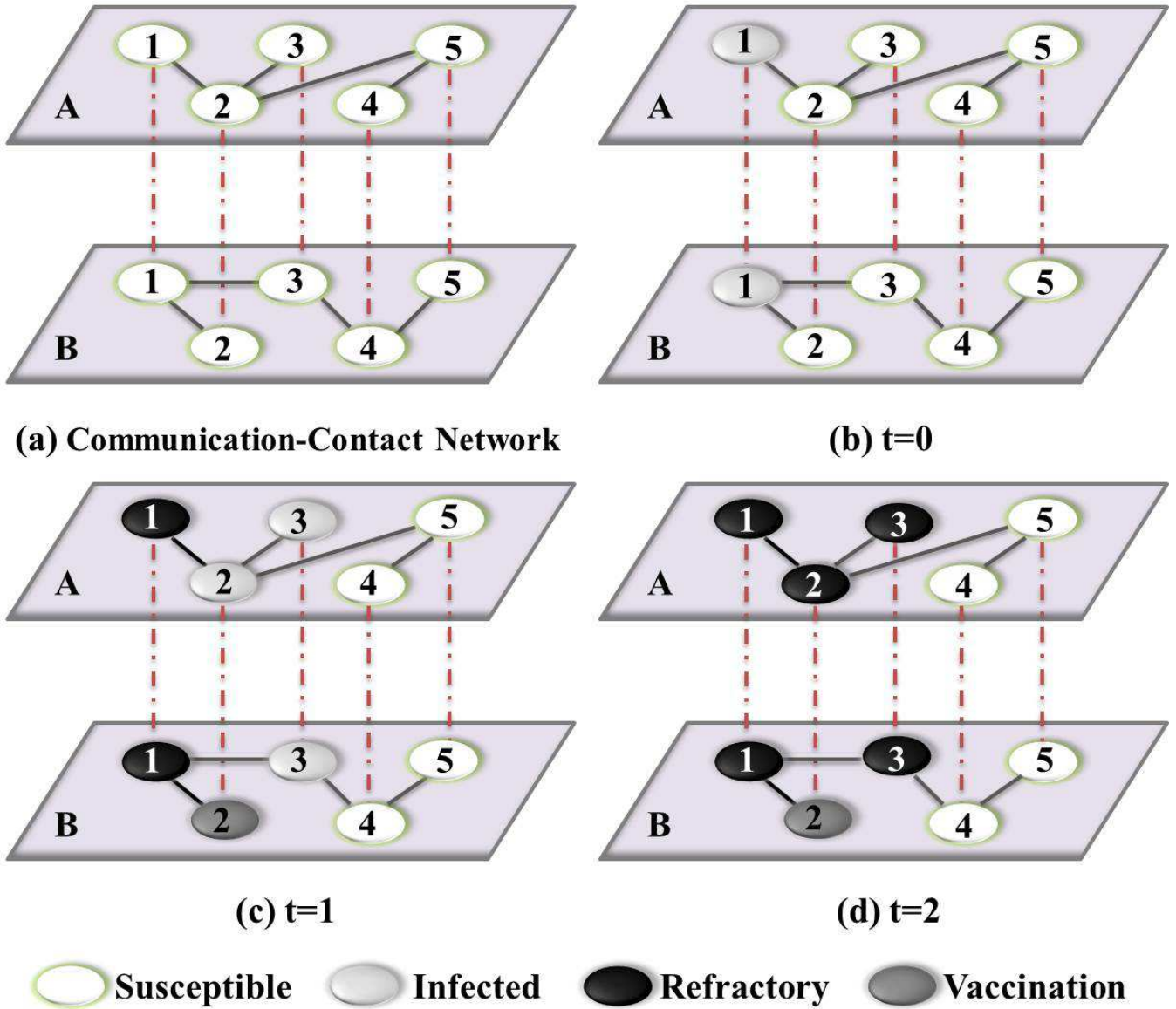,width=1\linewidth}
\caption{Illustration of asymmetrically coupled spreading processes on a
simulated communication-contact double-layer network. (a) Communication
and contact networks, denoted as layer $A$ and layer $B$, respectively,
each of five nodes. (b) At $t=0$, node $B_1$ in layer $B$ is randomly
selected as the initial infected node and its counterpart, node $A_1$
in layer $A$, gains the information that $B_1$ is infected, while all
other pairs of nodes, one from layer $A$ and another from layer $B$, are
in the susceptible state. (c) At $t=1$, within layer $A$ the information is
transmitted from $A_1$ to $A_2$ with probability $\beta_A$. Node $B_3$ in layer
$B$ can be infected by node $B_1$ with probability $\beta_B$ and, if it is indeed
infected, its corresponding node $A_3$ in layer $A$ gets the information as well.
Since, by this time, $A_2$ is already aware of the infection spreading, its
counterpart $B_2$ in layer $B$ is vaccinated, say with probability $p$. At the
same time, node $A_1$ in layer $A$ and its counterpart $B_1$ in layer $B$ enter
into the refractory state with probability $\mu_A$ and $\mu_B$, respectively.
(d) At $t=2$, all infected (or informed) nodes in both layers can
no longer infect others, and start recovering from the infection. In both
layers, the spreading dynamics terminate by this time.}
\label{fig:schematic}
\end{center}
\end{figure}

\textbf{\emph{Asymmetrically interacting spreading dynamics}.}
The dynamical processes of disease and information spreading are typically
asymmetrically coupled with each other. The dynamics component in our
model can be described, as follows. In the communication layer (layer $A$),
the classic susceptible-infected-recovered (SIR) epidemiological
model~\cite{Moreno:2002} is used to describe the dissemination of
information about the disease. In the SIR model, each node can be
in one of the three states: (1) susceptible state ($S$) in which the
individual has not received any information about the disease,
(2) informed state($I$), where the individual is aware of disease and
is capable of transmitting the information to other individuals in
the same layer, and (3) refractory state ($R$), in which the
individual has received the information but is not willing to pass
it on to other nodes. At each time step, the information can propagate
from every informed node to all its neighboring nodes. If a neighbor
is in the susceptible state, it will be informed with probability
$\beta_A$. At the same time, each informed node can enter the
recovering phase with probability $\mu_A$. Once an informed node is
recovered, it will remain in this state for all subsequent time. A
node in layer $A$ will get the information about the disease once its
counterpart node in layer $B$ is infected. As a result, dissemination
of the information over layer $A$ is facilitated by disease transmission
on layer $B$.

The spreading dynamics in layer $B$ can be described by the SIRV
model~\cite{Ruan:2012}, in which a fourth sate, the state of vaccination ($V$),
is introduced. Mathematically, the SIR component of the spreading dynamics
is identical to the dynamics on layer $A$ except for different infection
and recovery rates, denoted by $\beta_B$ and $\mu_B$, respectively. If a
node in layer $B$ is in the susceptible state but its counterpart node in
layer $A$ is in the infected state, the node in layer $B$ will be vaccinated
with probability $p$. Disease transmission in the contact layer can thus
be suppressed by dissemination of information in the communication layer.
The two spreading processes and their dynamical interplay are illustrated
schematically in Fig.~\ref{fig:schematic}. Without loss of generality, we
set $\mu_A=\mu_B=1$.

\textbf{\emph{Theory of spreading dynamics in uncorrelated double-layer
networks}.} Two key quantities in the dynamics of spreading are the
outbreak threshold and the fraction of infected nodes in the final steady
state. We develop a theory to predict these quantities for both information
and epidemic spreading in the double-layer network. In particular, we adopt
the heterogeneous mean-field theory~\cite{Barthelemy:2004} to uncorrelated
double-layer networks.

Let $P_A(k_A)$ and $P_B(k_B)$ be the degree distributions of layers $A$ and
$B$, with mean degree $\langle k_A\rangle$ and $\langle k_B\rangle$,
respectively. We assume that the subnetworks associated with both layers
are random with no degree correlation. The time evolution of the epidemic
spreading is described by the variables $s_{k_A}^A(t)$, $\rho_{k_A}^A(t)$,
and $r_{k_A}^A(t)$, which are the densities of the susceptible, informed,
and recovered nodes of degree $k_A$ in layer $A$ at time $t$, respectively.
Similarly, $s_{k_B}^B(t)$, $\rho_{k_B}^B(t), r_{k_B}^B(t)$, and $v_{k_B}^B(t)$
respectively denote the susceptible, infected, recovered, and vaccinated
densities of nodes of degree $k_B$ in layer $B$ at time $t$.

The mean-field rate equations of the information spreading in layer $A$ are
\begin{eqnarray}
\label{A1}
\frac{ds_{k_A}^A(t)}{dt} & = & -s_{k_A}^A(t)[\beta_{A}k_A\Theta_{A}(t)
 + \beta_{B}\Theta_{B}(t)\sum_{k_B}k_BP_{B}(k_B)], \\
\label{A2}
\frac{d\rho_{k_A}^{A}(t)}{dt} & = & s_{k_A}^A(t)[\beta_{A}k_A\Theta_{A}(t)
+\beta_{B}\Theta_{B}(t)\sum_{k_B}k_BP_{B}(k_B)] - \rho_{k_A}^A(t), \\
\label{A3}
\frac{dr_{k_A}^A(t)}{dt} & = & \rho_{k_A}^A(t).
\end{eqnarray}
The mean-field rate equations of epidemic spreading in layer $B$ are given by
\begin{eqnarray}
\label{B1}
\frac{ds_{k_B}^B(t)}{dt} & = & -s_{k_B}^B(t)\beta_{B}k_B\Theta_{B}(t)
 - p\beta_{A}\Theta_{A}(t)\sum_{k_A}s_{k_A}^A(t)k_AP_{A}(k_A), \\
\label{B2}
\frac{d\rho_{k_B}^{B}(t)}{dt} & = & s_{k_B}^B(t)\beta_{B}k_B\Theta_{B}(t)
-\rho_{k_B}^{B}(t), \\
\label{B3}
\frac{dr_{k_B}^{B}(t)}{dt} & = & \rho_{k_B}^{B}(t), \\
\label{B4}
\frac{dv_{k_B}^{B}(t)}{dt} & = & p\beta_{A}\Theta_{A}(t)\sum_{k_A}s_{k_A}^A(t)
k_AP_{A}(k_A),
\end{eqnarray}
where $\Theta_{A}(t)$ ($\Theta_{B}(t)$) is the probability that a neighboring
node in layer A (layer B) is in the informed (infected) state (See
{\bf Methods} for details).

From Eqs.~(\ref{A1})-(\ref{B4}), the density associated with each
distinct state in layer $A$ or $B$ is given by
\begin{equation} \label{final density}
X_{h}(t) = \sum_{k_{h}=1}^{k_{h,max}}P_{h}(k_h)X_{k_h}^{h}(t),
\end{equation}
where $h\in\{A,B\}$, $X\in\{S,I,R,V\}$, and $k_{h,max}$ denotes the largest
degree of layer $h$. The final densities of the whole system can be obtained
by taking the limit $t\rightarrow\infty$.

Due to the complicated interaction between the disease and information spreading
processes, it is not feasible to derive the exact threshold values. We
resort to a linear approximation method to get the outbreak threshold of
information spreading in layer $A$ (see Supporting Information for details) as
\begin{equation} \label{Athreshold}
\beta_{Ac} = \left\{\begin{array}{l}\beta_{Au},
\quad for \quad \beta_{B}\leq \beta_{Bu} \\
0, \quad \quad for  \quad \beta_{B}> \beta_{Bu},
\end{array} \right.
\end{equation}
where
\begin{eqnarray}
\nonumber
\beta_{Au} & \equiv & \langle k_A\rangle/({\langle k_A^2\rangle
-\langle k_A\rangle}) \ \ \mbox{and} \ \ \\ \nonumber
\beta_{Bu} & \equiv & \langle k_B\rangle/(\langle k_B^2\rangle-\langle k_B\rangle)
\end{eqnarray}
denote the outbreak threshold of information spreading in layer $A$ when it
is isolated from layer $B$, and that of epidemic spreading in layer $B$ when
the coupling between the two layers is absent, respectively.

Equation~(\ref{Athreshold}) has embedded within it two distinct physical
mechanisms for information outbreak. The first is the intrinsic information
spreading process on the isolated layer $A$ without the impact of the spreading
dynamics from layer $B$. For $\beta_{B}> \beta_{Bu}$, the outbreak of epidemic
will make a large number of nodes in layer $A$ ``infected'' with the information,
even if on layer $A$, the information itself cannot spread through the
population efficiently. In this case, the information outbreak has little
effect on the epidemic spreading in layer $B$ because very few nodes in this
layer are vaccinated. We thus have $\beta_{Bc}\approx\beta_{Bu}$ for
$\beta_{A}\leq\beta_{Au}$.

However, for $\beta_{A}>\beta_{Au}$, epidemic spreading in layer $B$ is
restrained by information spread, as the informed nodes in layer $A$ tend
to make their counterpart nodes in layer $B$ vaccinated. Once a node is in
the vaccination state, it will no longer be infected. In a general sense,
vaccination can be regarded as a type of ``disease,'' as every node
in layer $B$ can be in one of the two states: infected or vaccinated.
Epidemic spreading and vaccination can thus be viewed as a pair of competing
``diseases'' spreading in layer $B$~\cite{Karrer:2011}. As pointed out by
Karrer and Newman~\cite{Karrer:2011}, in the limit of large network size $N$,
the two competing diseases can be treated as if they were in fact spreading
non-concurrently, one after the other.

Initially, both epidemic and vaccination spreading processes exhibit
exponential growth (see Supporting Information). We can thus
obtain the ratio of their growth rates as
\begin{equation} \label{ratio}
\theta=\frac{\beta_B\beta_{Au}}{\beta_A\beta_{Bu}}.
\end{equation}
For $\theta>1$, the epidemic disease spreads faster than the vaccination.
In this case, the vaccination spread is insignificant and can be neglected.
For $\theta<1$, information spreads much faster than the disease, in
accordance with the situation in a modern society. Given that the
vaccination and epidemic processes can be treated successively and
separately, the epidemic outbreak threshold can be derived by a bond
percolation analysis~\cite{Newman:2005,Karrer:2011} (see details in Supporting
Information). We obtain
\begin{equation} \label{betaBc}
\beta_{Bc} = \frac{\langle k_{B}\rangle}{(1-pS_A)
(\langle k_{B}^{2}\rangle-\langle k_{B}\rangle)},
\end{equation}
where $S_A$ is the density of the informed population, which can be
obtained by solving Eqs.~(S18) and (S19) in Supporting Information.
For $\theta<1$, we see from Eq.~(\ref{betaBc}) that the threshold for
epidemic outbreak can be enhanced by the following factors: strong
heterogeneity in the communication layer, large information-transmission
rate, and large vaccination rate.

\textbf{\emph{Simulation results for uncorrelated networks}.}
We use the standard configuration model to generate networks with power-law
degree distributions~\cite{Newman:2005-2,Newman:2001-2,Catanzaro:2005} for the
communication subnetwork (layer A). The contact subnetwork in layer $B$
is of the Erd\H{o}s and R\'{e}nyi (ER) type~\cite{Erdos:1959}. We
use the notation SF-ER to denote the double-layer network. The sizes of both
layers are set to be $N_A=N_B=2 \times 10^4$ and their average degrees are
$\langle k_A\rangle=\langle k_B\rangle=8$. The degree distribution of the
communication layer is $P_A(k_A)= \zeta k_A^{-{\gamma_A}}$ with the coefficient
$\zeta=1/\sum_{k_{min}}^{k_{max}}k_A^{-{\gamma_A}}$ and the maximum
degree $k_{max} \sim N^{1/({\gamma_A}-1)}$. We focus on the case of
${\gamma_A}=3.0$ here in the main text (the results for other values
of the exponent, e.g., ${\gamma_A}=2.7$ and $3.5$, are similar, which
are presented in Supporting Information). The degree distribution of the
contact layer is $P_B(k_B)=e^{-\langle k_B\rangle} \langle k_B\rangle^{k_B}/k_B!$.
To initiate an epidemic spreading process, a node in layer $B$ is randomly
infected and its counterpart node in layer $A$ is thus in the informed state,
too. We implement the updating process with parallel dynamics, which is
widely used in statistical physics~\cite{Marro:1999} (see Sec. S3A in
Supporting Information for more details). The spreading dynamics
terminates when all infected nodes in both layers are recovered, and
the final densities $R_A$, $R_B$, and $V_B$ are then recorded.

\begin{figure}
\begin{center}
\epsfig{file=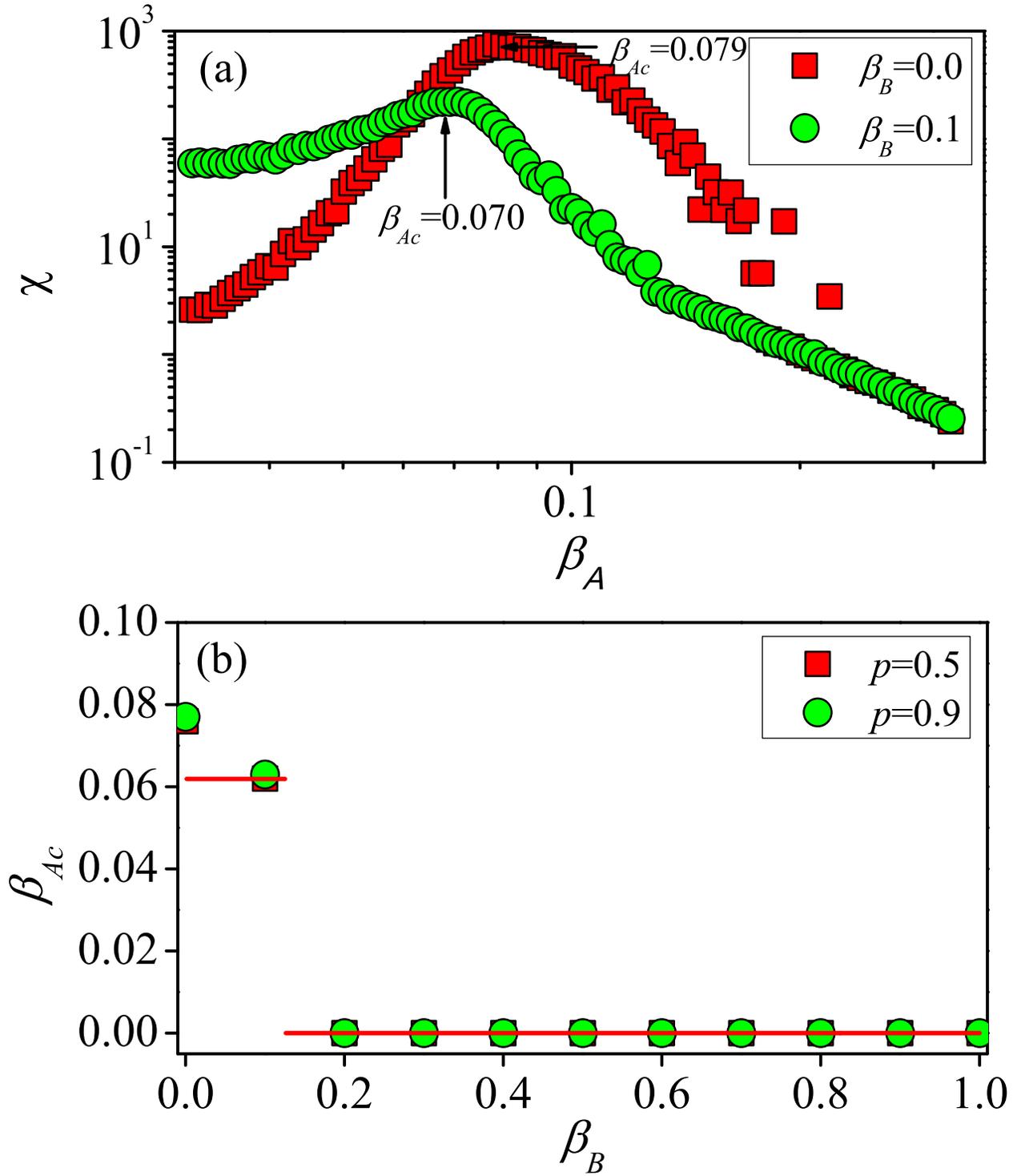,width=1\linewidth}
\caption{On SF-ER networks, (a) susceptibility measure
$\chi$ as a function of the information-transmission rate $\beta_A$ for
$p = 0.5$, $\beta_B=0.0$ (red squares) and $\beta_B=0.1$ (green circles),
(b) threshold $\beta_{Ac}$ of information spreading as a function
of the disease-transmission rate $\beta_B$ for vaccination rate $p=0.5$
(red squares) and $p=0.9$ (green circles), where the red solid lines are
analytical predictions from Eq.~(\ref{Athreshold}).}
\label{fig2}
\end{center}
\end{figure}

For epidemiological models [\emph{e.g.}, the susceptible-infected-susceptible
(SIS) and SIR] on networks with a power-law degree distribution, the
finite-size scaling method may not be effective to determine the critical point
of epidemic dynamics~\cite{Ferreira:2012,Hong:2007}, because the outbreak
threshold depends on network size and it goes to zero in the thermodynamic
limit~\cite{Boguna:2013,Moreno:2002}. Therefore, we employ the
\emph{susceptibility measure}~\cite{Ferreira:2012} $\chi$ to numerically
determine the size-dependent outbreak threshold:
\begin{equation}\label{chi}
\chi=N\frac{\langle r^{2}\rangle-\langle r\rangle^{2}}{\langle r\rangle},
\end{equation}
where $N$ is network size ($N = N_A=N_B$), and $r$ denotes the final outbreak
ratio such as the final densities $R_A$ and $R_B$ of the recovered nodes
in layers $A$ and $B$, respectively. We use $2\times10^3$ independent
dynamic realizations on a fixed double-layer network to calculate the
average value of $\chi$ for the communication layer for each value of
$\beta_A$. As shown in Fig.~\ref{fig2}(a), $\chi$ exhibits a maximum
value at $\beta_{Ac}$, which is the threshold value of the information
spreading process. The simulations are further implemented using $30$
different two-layer network realizations to obtain the average value
of $\beta_{Ac}$. The identical simulation setting is used for all
subsequent numerical results, unless otherwise specified. Figure~\ref{fig2}(b)
shows the information threshold $\beta_{Ac}$ as a function of the
disease-transmission rate $\beta_B$. Note that the statistical errors
are not visible here (same for similar figures in the paper), as they
are typically vanishingly small. We see that the behavior of the information
threshold can be classified into two classes, as predicted by
Eq.~(\ref{Athreshold}). In particular, for
$\beta_B\leq\beta_{Bu}=1/\langle k_B\rangle=0.125$,
the disease transmission on layer $B$ has little impact on the
information threshold on layer $A$, as we have
$\beta_{Ac}\approx\beta_{Au}= \langle k_A\rangle/({\langle k_A^2\rangle -
\langle k_A\rangle})\approx0.06$.
For $\beta_B>\beta_{Bu}$, the outbreak of epidemic on layer $B$ leads to
$\beta_{Ac}=0.0$. Comparison of the information thresholds for different
vaccination rates shows that the value of the vaccination probability $p$
has essentially no effect on $\beta_{Ac}$.

\begin{figure}
\begin{center}
\epsfig{file=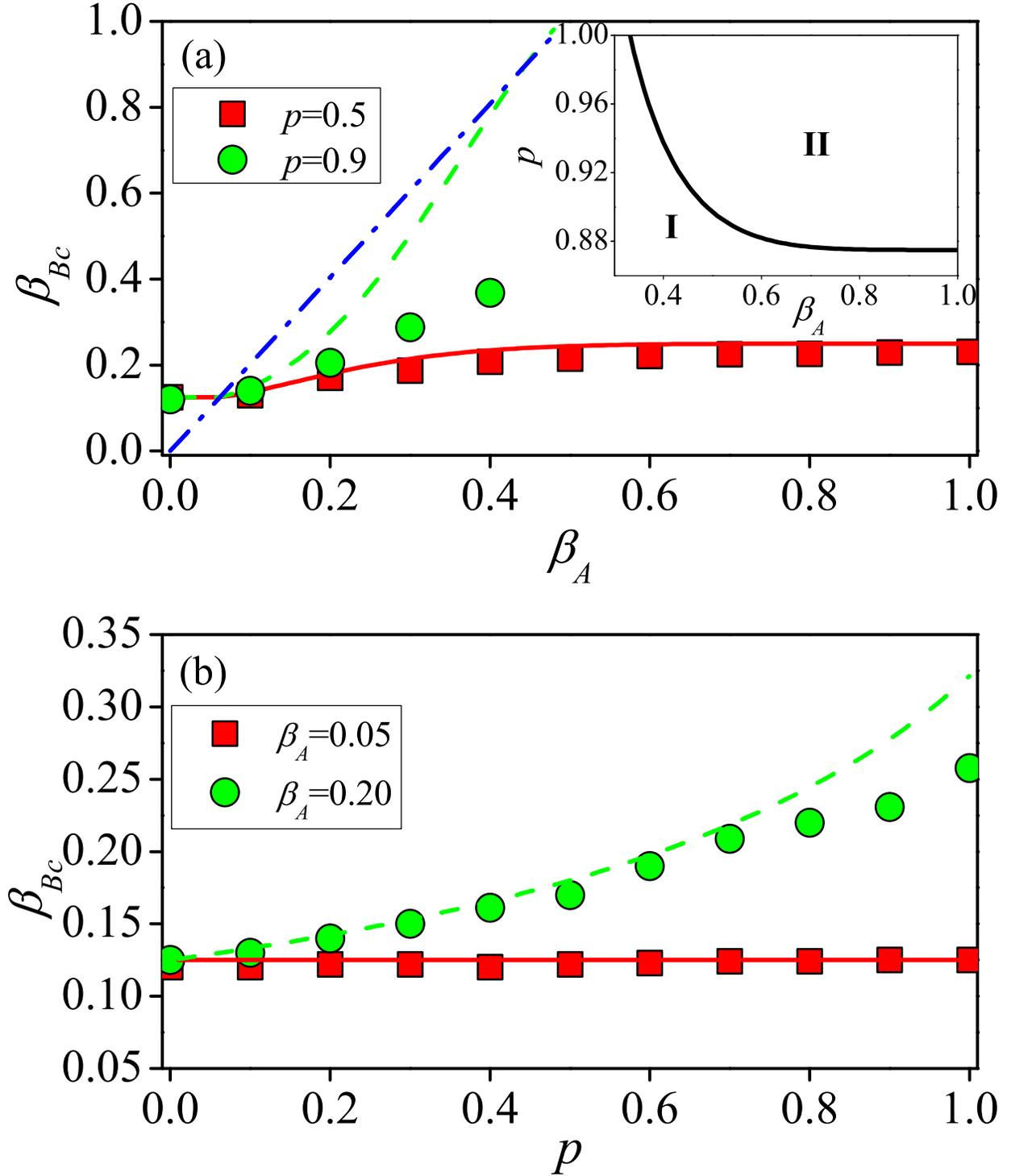,width=1\linewidth}
\caption{For SF-ER double-layer networks, epidemic threshold
$\beta_{Bc}$ as a function of the information-transmission rate $\beta_A$ (a)
and the vaccination rate $p$ (b). In (a), the red solid ($p=0.5$) and
green dashed ($p=0.9$) lines are the analytical predictions from
Eq.~(\ref{betaBc}), and the blue dot-dashed line denotes the case
of $\theta=1$ from Eq.~(\ref{ratio}). Iset of (a) shows the condition
under which a giant residual cluster of layer $B$ exists [from Eq.~(S20)
in Supporting Information] in phase I. In (b), the red solid line
($\beta_A=0.05$) corresponds to $\beta_{Bc}=\beta_{Bu}$, and the
green dashed line ($\beta_A=0.20$) is the analytical prediction
from Eq.~(\ref{betaBc}).}
\label{fig3}
\end{center}
\end{figure}

Figure~\ref{fig3} shows the effect of the information-transmission rate $\beta_A$
and the vaccination rate $p$ on the epidemic threshold $\beta_{Bc}$. From
Fig.~\ref{fig3}(a), we see that the value of $\beta_{Bc}$ is not influenced
by $\beta_A$ for $\beta_A\leq\beta_{Au}\approx0.06$, whereas
$\beta_{Bc}$ increases with $\beta_A$. For $p=0.5$, the analytical
results from Eq.~(\ref{betaBc}) are consistent with the simulated
results. However, deviations occur for larger values of $p$, e.g., $p=0.9$,
because the effect of information spreading is over-emphasized in
cases where the two types of spreading dynamics are treated
successively but not simultaneously. The gap between the theoretical
and simulated thresholds diminishes as the network size is increased,
validating applicability of the analysis method that, strictly speaking,
holds only in the thermodynamic limit~\cite{Karrer:2011} (see details in
Supporting Information). Note that a giant residual cluster does not
exist in layer $B$ for $p=0.9$ and $\beta_A\geq0.49$, ruling out
epidemic outbreak. The phase diagram indicating the possible existence of
a giant residual cluster [Eq.~(S20) in Supporting Information] is shown in the
inset of Fig.~\ref{fig3}(a), where in phase II, there is no such cluster.
In Fig.~\ref{fig3}(b), a large value of $p$ causes $\beta_{Bc}$ to increase
for $\beta_A>\beta_{Au}$. We observe that, similar to Fig.~\ref{fig3}(a),
for relatively large values of $p$, say $p \geq 0.8$, the analytical
prediction deviates from the numerical results. The effects of network
size $N$, exponent ${\gamma_A}$ and SF-SF network structure on the
information and epidemic thresholds are discussed in detail
in Supporting Information.

\begin{figure}
\begin{center}
\epsfig{file=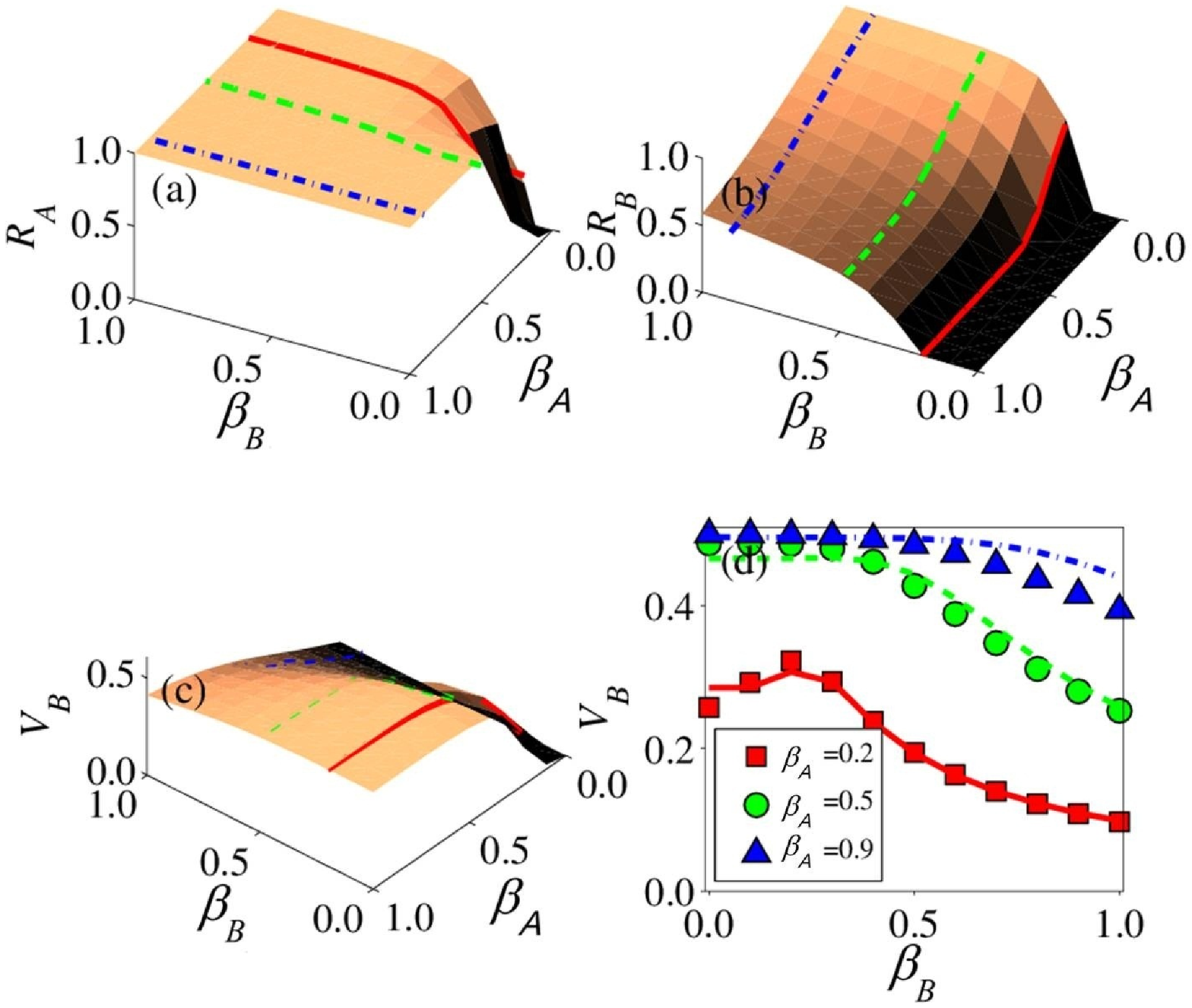,width=1\linewidth}
\caption{For SF-ER networks, final density in each state
versus the parameters $\beta_A$ and $\beta_B$: (a) recovered density $R_A$,
(b) recovered density $R_B$, (c) the vaccination density $V_B$, and (d)
$V_B$ versus $\beta_B$ for $\beta_A=0.2,0.5,0.9$. The value of parameter $p$
is 0.5. Different lines are the numerical solutions of
Eqs.~(\ref{A1})-(\ref{final density}) in the limit $t\rightarrow\infty$.
In (a) and (d), we select three different values of $\beta_A$ (0.2, 0.5, and
0.9), corresponding to the red solid, green dashed, and blue dot-dashed lines,
respectively. In (b) and (c), three different values of $\beta_B$ are chosen
(0.2, 0.5, and 0.9), corresponding to the red solid, green dashed, and blue
dot-dashed lines, respectively.}
\label{fig4}
\end{center}
\end{figure}

The final dynamical state of the double-layer spreading system is shown
in Fig.~\ref{fig4}. From Fig.~\ref{fig4}(a), we see that the final
recovered density $R_A$ for information increases with $\beta_{A}$ and
$\beta_{B}$ rapidly for $\beta_{A}\leq\beta_{Au}$
and $\beta_{B}\leq\beta_{Bu}$. Figure~\ref{fig4}(b) reveals that the recovered
density $R_B$ for disease decreases with $\beta_{A}$. We see that a large
value of $\beta_{A}$ can prevent the outbreak of epidemic for small
values of $\beta_{B}$, as $R_B \rightarrow 0$ for $\beta_{B}=0.2$ and
$\beta_{A} \geq 0.5$ (the red solid line). From Fig.~\ref{fig4}(c),
we see that, with the increase in $\beta_{A}$, more nodes in layer $B$
are vaccinated. It is interesting to note that the vaccinated density
$V_B$ exhibits a maximum value if $\beta_{A}$ is not large. Figure~\ref{fig4}
shows that the maximum value of $V_B$ is about $0.32$, which occurs
at $\beta_{B} \approx 0.20$, for $\beta_{A}=0.2$. Combining with
Fig.~\ref{fig3}(a), we find that the corresponding point of the
maximum value $\beta_{B} \approx 0.20$ is close to $\beta_{Bc} \approx 0.16$
for $p=0.5$. This is because the transmission of disease has the opposite
effects on the vaccinations. For $\beta_{B}\leq\beta_{Bc}$, the newly
infected nodes in layer $B$ will facilitate information spreading in
layer $A$, resulting in more vaccinated nodes. For $\beta_{B}>\beta_{Bc}$,
the epidemic spreading will make a large number of nodes infected, reducing
the number of nodes that are potentially to be vaccinated. For relatively
large values of $\beta_{A}$, information tends to spread much faster than
the disease for $\beta_{B}\leq\beta_{Bc}$, e.g.,
$\theta \approx 0.21$ for $\beta_{A}=0.5$, $p=0.5$, $\beta_{Bc}\approx 0.22$,
and $\theta \approx 0.12$ for $\beta_{A}=0.9$, $p=0.5$, and
$\beta_{Bc}\approx0.23$. In this case, the effect of disease transmission on
information spreading is negligible. The densities of the final dynamical
states for SF-SF networks are also shown in Supporting Information, and
we observe similar behaviors.

\begin{figure}
\begin{center}
\epsfig{file=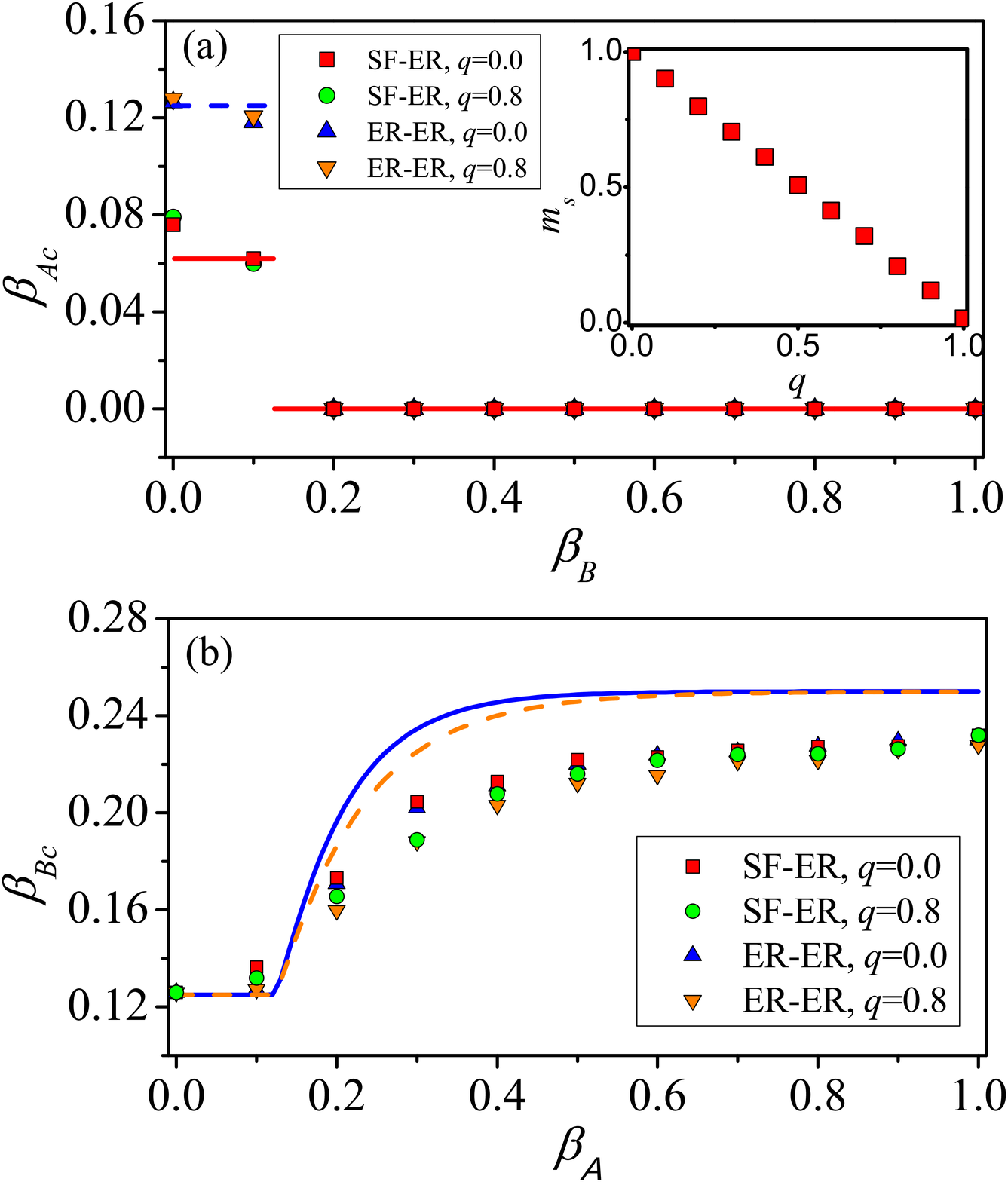,width=1\linewidth}
\caption{For two-layer correlated networks with vaccination
probability $p = 0.5$, the effect of one type of spreading dynamics
on the outbreak threshold of the counter-type spreading dynamics.
(a) $\beta_{Ac}$ versus $\beta_{B}$ on SF-ER networks with $q=0.0$ (red squares)
and $q=0.8$ (green circles), and ER-ER networks with $q=0.0$ (blue up triangles)
and $q=0.8$ (orange down triangles). Red solid (SF-ER) and blue dashed (ER-ER)
lines are the analytical predictions from Eq.~(9) and Eq.~(S37)
(in Supporting Information), respectively.
The inset shows the inter-layer correlation $m_s$ as a function of rematching
probability $q$. (b) $\beta_{Bc}$ versus $\beta_{A}$ on SF-ER networks with
$q=0.0$ (red squares) and $q=0.8$ (green circles), and ER-ER networks with
$q=0.0$ (blue up triangles) and $q=0.8$ (orange down triangles). Blue
solid ($q=0.0$) and orange dashed ($q=0.8$) lines are the analytical predictions
for ER-ER networks from Eqs.~(S38)-(S41) in Supporting Information.}
\label{fig5}
\end{center}
\end{figure}

\textbf{\emph{Spreading dynamics on correlated double-layer networks}.}
In realistic multiplex networks certain degree of inter-layer correlations
is expected to exist~\cite{Kivela:2013}. For example, in social networks,
positive inter-layer correlation is more common than negative
correlation~\cite{Szell:2010,Nicosia:2013}. That is,
an ``important'' individual with a large number of links in one network layer
(e.g., representing one type of social relations) tends to have many links
in other types of network layers that reflect different kinds of social
relations. Recent works have shown that inter-layer correlation can have
a large impact on the percolation properties of multiplex
networks~\cite{Parshani:2010,Lee:2012}. Here, we investigate how the
correlation between the communication and contact layers affects the
information and disease spreading dynamics. To be concrete, we focus
on the effects of positive correlation on the two types of spreading
dynamics. It is necessary to construct a two-layer correlated network
with adjustable degree of inter-layer correlation. This can be accomplished
by first generating a two-layer network with the maximal positive
correlation, where each layer has the same structure as uncorrelated
networks. Then, $Nq$ pairs of counterpart nodes, in which $q$ is the
rematching probability, are rematched randomly, leading to a two-layer
network with weaker inter-layer correlation. The inter-layer correlation
after rematching is given by (see {\bf Methods})
\begin{equation}\label{rsq}
m_s \approx 1-q,
\end{equation}
which is consistent with the numerical results [e.g., see inset of
Fig.~\ref{fig5}(a) below]. For SF-ER networks with fixed correlation
coefficient, the mean-field rate equations of the double-layer system
cannot be written down because the concrete expressions of the
conditional probabilities $P(k_B|k_A)$ and $P(k_A|k_B)$ are no
longer available.

\begin{figure}
\begin{center}
\epsfig{file=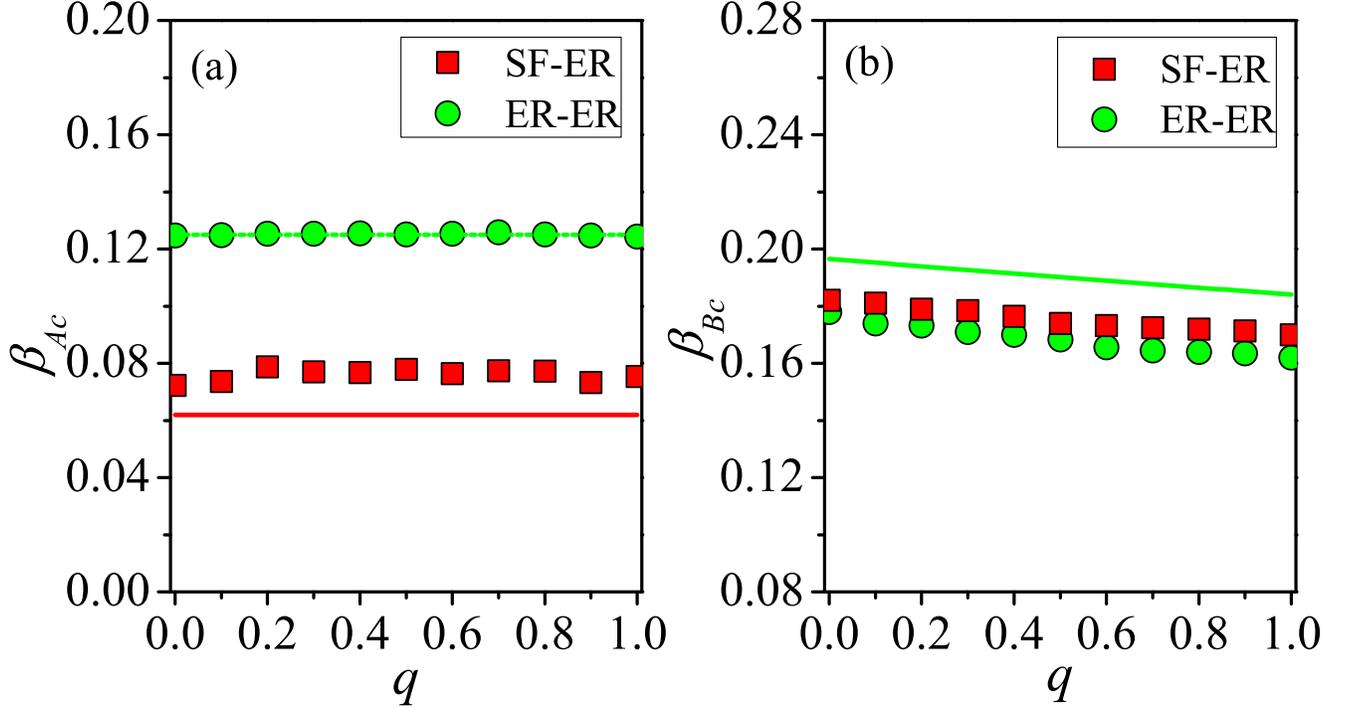,width=1\linewidth}
\caption{Effect of varying the rematching probability on outbreak
thresholds of the two types of spreading dynamics. (a) $\beta_{Ac}$ versus $q$
on SF-ER (red squares) and ER-ER networks (green circles) for $\beta_{B}=0.05$
and $p=0.5$. Red Solid (SF-ER) and green dashed (ER-ER) lines are analytical
predictions from Eq.~(9) and Eq.~(S37) in Supporting Information, respectively.
(b) $\beta_{Bc}$ versus $q$ on SF-ER (red squares) and ER-ER networks
(green circles) for $\beta_{A}=0.2$ and $p=0.5$. Green solid line is
analytical prediction for ER-ER networks from Eqs.~(S38)-(S41) in
Supporting Information.}
\label{fig6}
\end{center}
\end{figure}

\begin{figure}
\begin{center}
\epsfig{file=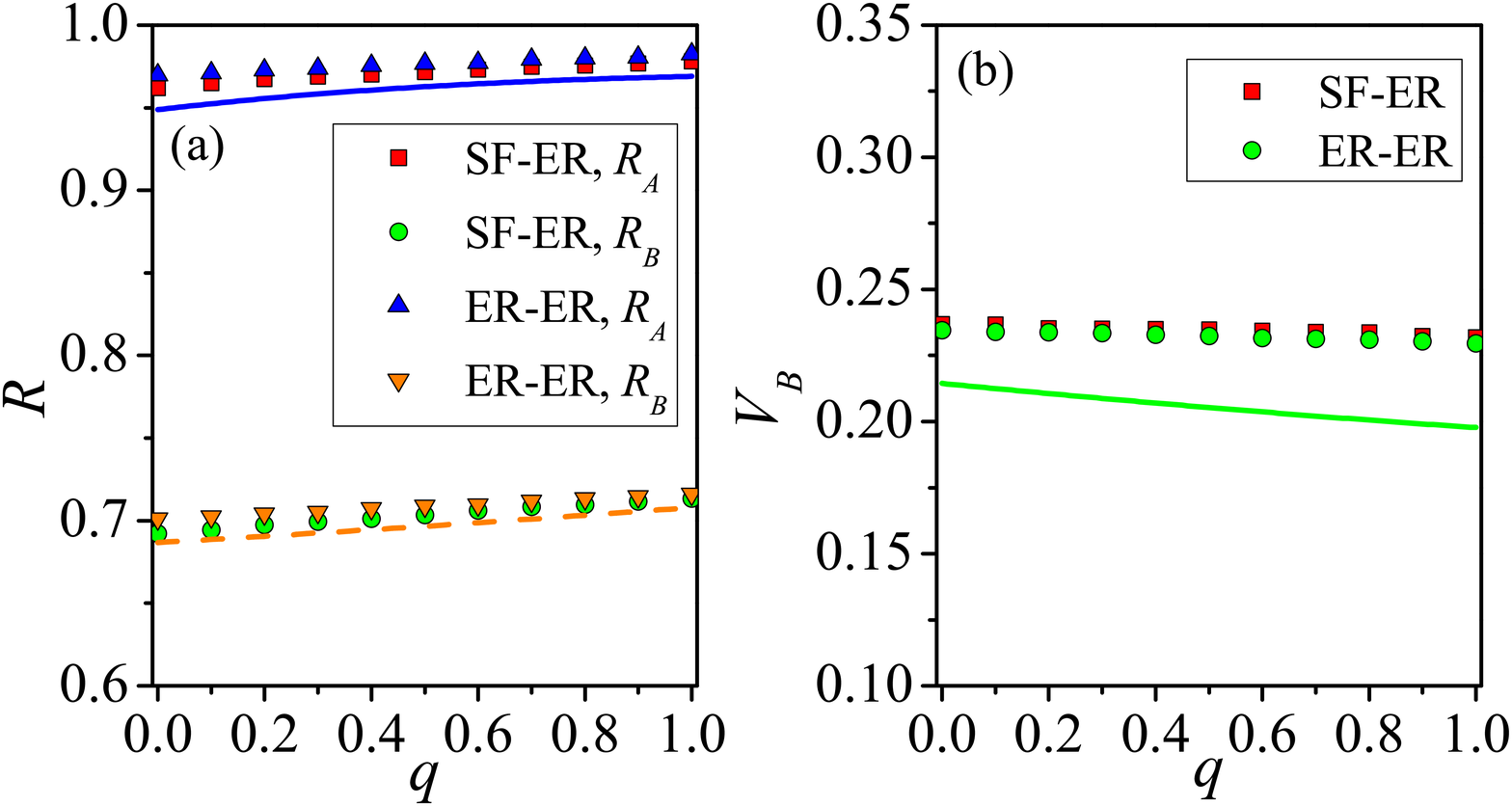,width=1\linewidth}
\caption{Effect of rematching probability on the final state.
(a) $R_{A}$ versus $q$ on SF-ER (red squares) and ER-ER networks (blue
up triangles), $R_{B}$ versus $q$ on SF-ER (green circles) and ER-ER
networks (orange down triangles). (b) $V_{B}$ versus $q$ on SF-ER (red
squares) and ER-ER networks (green circles). Different lines represent
the analytic solutions for ER-ER networks, calculated by summing the
final densities of all degrees from Eqs.~(S28)-(S34) in Supporting
Information. The parameter setting is $\beta_{A}=0.2$, $\beta_{B}=0.4$
and $p=0.5$.}
\label{fig7}
\end{center}
\end{figure}

We investigate how the rematching probability $q$ affects the outbreak
thresholds in both the communication and epidemic layers. As shown in
Fig.~\ref{fig5}, we compare the case of $q=0.8$ with that of $q=0.0$.
From Fig.~\ref{fig5}(a), we see that $q$ has little impact on the outbreak
threshold $\beta_{Ac}$ of the communication layer [with further support
in Fig.~\ref{fig6}(a), and analytic explanation using ER-ER correlated
layered networks in Supporting Information]. We also see that the
value of $\beta_{Ac}$ for ER-ER layered networks with the same mean degree
is greater because of the homogeneity in the degree distribution of layer $A$.
Figures~\ref{fig5}(b) and~\ref{fig6}(b) show that $\beta_{Bc}$ decreases with
$q$ or, equivalently, $\beta_{Bc}$ increases with $m_s$. This is because
stronger inter-layer correlation can increase the probability for nodes with
large degrees in layer $B$ to be vaccinated, thus effectively preventing the
outbreak of epidemic [see also Eqs.~(S38)-(S41) in Supporting Information].
Figure~\ref{fig7} shows the final densities of different populations,
providing the consistent result that, with the increase (decrease) of
$q$ ($m_s$), the final densities $R_A$ and $R_B$ increase but the density
$V_B$ decreases. For SF-SF networks, we obtain similar results
(shown in Supporting Information).

\section*{Discussion}

To summarize, we have proposed an asymmetrically interacting, double-layer
network model to elucidate the interplay between information diffusion
and epidemic spreading, where the former occurs on one layer (the communication
layer) and the latter on the counterpart layer. A mean-field
based analysis and extensive computations reveal an intricate interdependence
of two basic quantities characterizing the spreading dynamics on both layers:
the outbreak thresholds and the final fractions of infected nodes. In particular,
on the communication layer, the outbreak of the information about the disease
can be triggered not only by its own spreading dynamics but also by the the
epidemic outbreak on the counter-layer. In addition, high disease
and information-transmission rates can enhance markedly the final
density of the informed or refractory population. On the layer of physical
contact, the epidemic threshold can be increased but only if information
itself spreads through the communication layer at a high rate.
The information spreading can greatly reduce the final refractory
density for the disease through vaccination. While a rapid spread
of information will prompt more nodes in the contact layer to consider
immunization, the authenticity of the information source must be
verified before administrating large-scale vaccination.

We have also studied the effect of inter-layer correlation on the
spreading dynamics, with the finding that stronger correlation has
no apparent effect on the information threshold, but it can suppress
the epidemic spreading through timely immunization of large-degree
nodes~\cite{Holme:2002}. These results indicate that it is possible
to effectively mitigate epidemic spreading through information diffusion,
e.g., by informing the high-centrality hubs about the disease.

The challenges of studying the intricate interplay between
social and biological contagions in human populations are
generating interesting science~\cite{Bauch:2013}. In this work, we study
asymmetrically interacting information-disease dynamics theoretically
and computationally, with implications to behavior-disease coupled
systems and articulation of potential epidemic-control strategies.
Our results would stimulate further works in the more realistic situation
of asymmetric interactions.

During the final writing of this paper, we noted one preprint posted online
studying the dynamical interplay between awareness and epidemic spreading
in multiplex networks~\cite{Granell:2013}. In that work, the two
competing infectious strains are described by two SIS processes.
The authors find that the epidemic threshold depends on the topological
structure of the multiplex network and the interrelation with the
awareness process by using a Markov-chain approach.
Our work thus provides further understanding and insights into
spreading dynamics on multi-layer coupled networks.

\section*{Methods}
\textbf{\emph{Mean-Field theory for the uncorrelated double-layer networks}.}
To derive the mean-field rate equations for the density variables, we
consider the probabilities that node $A_i$ in layer $A$ and node $B_i$
in layer $B$ become infected during the small time interval $[t, t+dt]$.
On the communication layer, a susceptible node $A_i$ of degree $k_A$ can
obtain the information in two ways: from its neighbors in the same
layer and from its counterpart node in layer $B$. For the first route,
the probability that node $A_i$ receives information from one of its neighbors
is $k_A\beta_A\Theta_A(t)dt$, where $\Theta_A(t)$ is the probability that
a neighboring node is in the informed state~\cite{Newman:2010} and is given by
\begin{equation}\label{Theta_A}
\Theta_{A}(t)=\frac{\sum_{k_A^{\prime}}(k_A^{\prime}-1)P_{A}(k_A^{\prime})
\rho_{k_{A}^{\prime}}^{A}(t)}{\langle k_{A}\rangle},
\end{equation}
where $\langle k_A\rangle=\sum_{k_A}k_AP_{A}(k_A)$.
To model the second scenario, we note that, due to the asymmetric coupling
between the two layers, a node in layer $A$ being susceptible requires that
its counterpart node in layer $B$ be susceptible, too. A node in the communication
layer will get the information about the disease once its counterpart node
in layer $B$ is infected, which occurs with the probability
$\sum_{k_B}P(k_B|k_A) k_B\beta_B\Theta_B(t)dt$, where $P(k_B|k_A)$ denotes
the conditional probability that a node of degree $k_A$ in layer $A$ is linked
to a node of degree $k_B$ in layer $B$, and $k_B\beta_B\Theta_B(t)dt$ is the
probability for a counterpart node of degree $k_B$ to become infected in the
time interval $[t,t+dt]$. If the subnetworks in both layers are not correlated,
we have $P(k_B|k_A)=P_B(k_B)$. The mean-field rate equations of the information
spreading in layer $A$ are Eqs.~(\ref{A1})-(\ref{A3}).

On layer $B$, a susceptible node $B_i$ of degree $k_B$ may become infected
or vaccinated in the time interval $[t,t+dt]$. This can occur in two ways.
Firstly, it may be infected by a neighboring node in the same layer with
the probability $k_B\beta_B\Theta_B(t)dt$, where $\Theta_B(t)$ is the
probability that a neighbor is in the infected state and is given by
\begin{equation} \label{Theta_B}
\Theta_{B}(t) = \frac{\sum_{k_B^{\prime}}(k_B^{\prime}-1)P_{B}(k_B^{\prime})
\rho_{k_{B}^{\prime}}^{B}(t)}{\langle k_{B}\rangle},
\end{equation}
where $\langle k_B\rangle=\sum_{k_B}k_BP_{B}(k_B)$. Secondly, if its
counterpart node in layer $A$ has already received the information from
one of its neighbors, it will be vaccinated with probability $p$. The
probability for a node in layer $B$ to be vaccinated, taking into account
the interaction between the two layers, is
$p\sum_{k_A}P(k_A|k_B)s_{k_A}^A(t)\beta_A k_A\Theta_A(t)dt$, where $P(k_A|k_B)$
denotes the conditional probability that a node of degree $k_B$ in layer $B$
is linked to a node of degree $k_A$ in layer $A$, and
$s_{k_A}^A(t)\beta_Ak_A\Theta_A(t)dt$ is the informed probability for the
counterpart node of degree $k_A$ in the susceptible state
[$P(k_A|k_B)=P_A(k_A)$ for $m_s=0$]. The mean-field rate equations of
epidemic spreading in layer $B$ are Eqs.~(\ref{B1})-(\ref{B4}).
We note that the second term on the right side of Eq.~(\ref{B1}) does not
contain the variable $s_{k_B}^B(t)$ because a node in layer $B$ must be in
the susceptible state if its counterpart node in layer $A$ is in the susceptible
state.

\textbf{\emph{Spearman rank correlation coefficient}}.
The correlation between the layers can be quantified by the Spearman rank
correlation coefficient~\cite{Cho:2010,Lee:2012} defined as
\begin{equation} \label{eq:Spearman_rank}
m_s=1-6\frac{\sum_{i=1}^{N}\Delta_i^{2}}{N(N^{2}-1)},
\end{equation}
where $N$ is network size and $\Delta_i$ denotes the difference between
node $i$'s degrees in the two layers. When a node in layer $A$ is matched
with a random node in layer $B$, $m_s$ is approximately zero in the
thermodynamic limit. In this case, the double-layer network is
uncorrelated~\cite{Cho:2010,Lee:2012}. When every node has the same rank of degree
in both layers, we have $m_s \approx 1$. In this case, there is a maximally
positive inter-layer correlation where, for example, the hub node with the
highest degree in layer $A$ is matched with the largest hub in layer $B$,
and the same holds for the nodes with the smallest degree. In the case of
maximally negative correlation, the largest hub in one layer is matched
with a node having the minimal degree in the other layer, so we have
$m_s \approx -1$.

In a double-layer network with the  maximally positive correlation, any pair of
nodes having the same rank of degree in the respective layers are matched, i.e.,
$\Delta_i=0$ for any pair of nodes $A_i$ and $B_i$. We thus have $m_s=1$,
according to Eq.~(\ref{eq:Spearman_rank}). After random rematching, a pair of
nodes have $\Delta_i=0$ with probability $1-q$ and a random difference $\Delta_i'$
with probability $q$. Equation~(\ref{eq:Spearman_rank}) can then be rewritten as
\begin{equation} \label{qrs}
m_s = 1-6\frac{q\sum_{i=1}^{N}\Delta_i'^{2}}{N(N^{2}-1)}.
\end{equation}
When all nodes are randomly rematched, the layers in the network
are completely uncorrelated, i.e., $m_s\approx0$. In this case, we have
\begin{equation}\label{qrsrandom}
6\frac{\sum_{i=1}^{N}\Delta_i'^{2}}{N(N^{2}-1)} \approx 1.
\end{equation}
Submitting Eq.~(\ref{qrsrandom}) into Eq.~(\ref{qrs}), the inter-layer
correlation after rematching is given by
\begin{equation}\label{rsq_1}
m_s \approx 1-q.
\end{equation}

\section*{Figure legends}

{\bf FIG 1}: Illustration of asymmetrically coupled
spreading processes on a simulated communication-contact double-layer
network. (a) Communication and contact networks,
denoted as layer $A$ and layer $B$, respectively, each of five nodes. (b) At $t=0$,
node $B_1$ in layer $B$ is randomly selected as the initial infected node and its
counterpart, node $A_1$ in layer $A$, gains the information that $B_1$ is infected,
while all other pairs of nodes, one from layer $A$ and another from layer $B$, are
in the susceptible state. (c) At $t=1$, within layer $A$ the information is
transmitted from $A_1$ to $A_2$ with probability $\beta_A$. Node $B_3$ in layer
$B$ can be infected by node $B_1$ with probability $\beta_B$ and, if it is indeed
infected, its corresponding node $A_3$ in layer $A$ gets the information as well.
Since, by this time, $A_2$ is already aware of the infection spreading, its
counterpart $B_2$ in layer $B$ is vaccinated, say with probability $p$. At the
same time, node $A_1$ in layer $A$ and its counterpart $B_1$ in layer $B$ enter
into the refractory state with probability $\mu_A$ and $\mu_B$, respectively.
(d) At $t=2$, all infected nodes in both layers can no longer infect others, and
start recovering from the infection. In both layers, the spreading dynamics
terminate by this time.

{\bf FIG 2}:
On SF-ER networks, (a) the susceptibility measure
$\chi$ as a function of the information-transmission rate $\beta_A$ for
$p = 0.5$, $\beta_B=0.0$ (red squares) and $\beta_B=0.1$ (green circles),
(b) the threshold $\beta_{Ac}$ of information spreading as a function
of the disease-transmission rate $\beta_B$ for vaccination rate $p=0.5$
(red squares) and $p=0.9$ (green circles), where the red solid lines are
analytical predictions from Eq.~(\ref{Athreshold}).

{\bf FIG 3}:
 For SF-ER double-layer networks, epidemic threshold
$\beta_{Bc}$ as a function of the information-transmission rate $\beta_A$ (a)
and the vaccination rate $p$ (b). In (a), the red solid ($p=0.5$) and
green dashed ($p=0.9$) lines are the analytical predictions from Eq.~(\ref{betaBc}),
and the blue dot-dashed line denotes the case of $\theta=1$ from Eq.~(\ref{ratio}).
The inset of (a) shows the condition under which a giant residual cluster of layer
$B$ exists [from Eq.~(S20) in Supporting Information] in phase
I. In (b), the red solid line
($\beta_A=0.05$) corresponds to $\beta_{Bc}=\beta_{Bu}$, and the green dashed line
($\beta_A=0.20$) is the analytical prediction from Eq.~(\ref{betaBc}).

{\bf FIG 4}: For SF-ER networks, the final density in each state
versus the parameters $\beta_A$ and $\beta_B$: (a) recovered density $R_A$,
(b) recovered density $R_B$, (c) the vaccination density $V_B$, and (d)
$V_B$ versus $\beta_B$ for $\beta_A=0.2,0.5,0.9$. The value of parameter $p$
is 0.5. Different lines are the numerical solutions of
Eqs.~(\ref{A1})-(\ref{final density}) in the limit $t\rightarrow\infty$.
In (a) and (d), we select three different values of $\beta_A$ (0.2, 0.5, and
0.9), corresponding to the red solid, green dashed, and blue dot-dashed lines,
respectively. In (b) and (c), three different values of $\beta_B$ are chosen
(0.2, 0.5, and 0.9), corresponding to the red solid, green dashed, and blue
dot-dashed lines, respectively.

{\bf FIG 5}: For two-layer correlated networks with vaccination
probability $p = 0.5$, the effect of one type of spreading dynamics
on the outbreak threshold of the counter-type spreading dynamics.
(a) $\beta_{Ac}$ versus $\beta_{B}$ on SF-ER networks with $q=0.0$ (red squares)
and $q=0.8$ (green circles), and ER-ER networks with $q=0.0$ (blue up triangles)
and $q=0.8$ (orange down triangles). Red solid (SF-ER) and blue dashed (ER-ER)
lines are the analytical predictions from Eq.~(9) and Eq.~(S37)
(in Supporting Information), respectively.
The inset shows the inter-layer correlation $m_s$ as a function of rematching
probability $q$. (b) $\beta_{Bc}$ versus $\beta_{A}$ on SF-ER networks with
$q=0.0$ (red squares) and $q=0.8$ (green circles), and ER-ER networks with
$q=0.0$ (blue up triangles) and $q=0.8$ (orange down triangles). Blue
solid ($q=0.0$) and orange dashed ($q=0.8$) lines are the analytical predictions
for ER-ER networks from Eqs.~(S38)-(S41) in Supporting Information.

{\bf FIG 6}: Effect of varying the rematching probability on outbreak
thresholds of the two types of spreading dynamics. (a) $\beta_{Ac}$ versus $q$
on SF-ER (red squares) and ER-ER networks (green circles) for $\beta_{B}=0.05$
and $p=0.5$. Red Solid (SF-ER) and green dashed (ER-ER) lines are analytical
predictions from Eq.~(9) and Eq.~(S37) in Supporting Information, respectively.
(b) $\beta_{Bc}$ versus $q$
on SF-ER (red squares) and ER-ER networks (green circles) for $\beta_{A}=0.2$
and $p=0.5$. Green solid line is analytical prediction for ER-ER networks from
Eqs.~(S38)-(S41) in Supporting Information.

{\bf FIG 7}: Effect of rematching probability on the final state.
(a) $R_{A}$ versus $q$ on SF-ER (red squares) and ER-ER networks (blue up triangles),
$R_{B}$ versus $q$ on SF-ER (green circles) and ER-ER networks (orange down triangles).
(b) $V_{B}$ versus $q$ on SF-ER (red squares) and ER-ER networks (green circles).
Different lines represent the analytic solutions for ER-ER networks, calculated by
summing the final densities of all degrees from Eqs.~(S28)-(S34) in Supporting
Information. The parameter setting is $\beta_{A}=0.2$, $\beta_{B}=0.4$ and $p=0.5$.\\

\section*{Acknowledgement}
M.T. would like to thank Prof. Pakming Hui for stimulating discussions.
This work was partially supported by the National Natural Science Foundation of
China (Grant No.~11105025) and China Postdoctoral Science Special Foundation (Grant No. 2012T50711).
Y. Do was supported by Basic Science Research Program through
the National Research Foundation of Korea (NRF) funded by the Ministry of Education,
Science and Technology (NRF-2013R1A1A2010067). Y.C.L. was
supported by AFOSR under Grant No. FA9550-10-1-0083.
GW Lee was supported by the Korea Meteorological Administration Research and
Development Program under Grant CATER 2012-2072.

\section*{Author contributions}
W. W., M. T. and Y. C. L devised the research project.
W. W. and H. Y. performed numerical simulations.
W. W., M. T., Y. H. D. and Y. C. L analyzed the results.
W. W., M. T., Y. H. D., Y. C. L and GW. L wrote the paper.

\section*{Additional information}



{\bf Competing financial interests}:
The authors declare no competing financial interests.

\newpage

\setcounter{page}{1}

\noindent${\Large\textbf{S1. Spreading dynamics on uncorrelated
doubule-layer networks
}}$\\

We adopt the heterogeneous mean-field theory~\cite{Barthelemy11:2004} to uncorrelated
double-layer networks. Let $P_A(k_A)$ and $P_B(k_B)$ be the degree distributions of layers $A$ and
$B$, with mean degree $\langle k_A\rangle$ and $\langle k_B\rangle$,
respectively. We assume that the subnetworks associated with both layers
are random with no degree correlation. The time evolution of the epidemic
spreading is described by the variables $s_{k_A}^A(t)$, $\rho_{k_A}^A(t)$,
and $r_{k_A}^A(t)$, which are the densities of the susceptible, infected,
and recovered nodes of degree $k_A$ in layer $A$ at time $t$, respectively.
Similarly, $s_{k_B}^B(t)$, $\rho_{k_B}^B(t), r_{k_B}^B(t)$, and $v_{k_B}^B(t)$
respectively denote the susceptible, infected, recovered, and vaccinated
densities of nodes of degree $k_B$ in layer $B$ at time $t$.\\

\noindent${\Large\textbf{A. Mean-field rate equations}}$\\

The mean-field rate equations of the information
spreading in layer $A$ are then
$$\frac{ds_{k_A}^A(t)}{dt}=-s_{k_A}^A(t)[\beta_{A}k_A\Theta_{A}(t)+ \beta_{B}\Theta_{B}(t)\sum_{k_B}k_BP_{B}(k_B)],  \eqno (\text{S}1) $$

$$\frac{d\rho_{k_A}^{A}(t)}{dt} = s_{k_A}^A(t)[\beta_{A}k_A\Theta_{A}(t)+\beta_{B}\Theta_{B}(t)\sum_{k_B}k_BP_{B}(k_B)] - \rho_{k_A}^A(t), \eqno (\text{S}2) $$

$$\frac{dr_{k_A}^A(t)}{dt} = \rho_{k_A}^A(t). \eqno (\text{S}3)$$

The mean-field rate equations of
epidemic spreading in layer $B$ are thus given by

$$\frac{ds_{k_B}^B(t)}{dt}  =  -s_{k_B}^B(t)\beta_{B}k_B\Theta_{B}(t)
- p\beta_{A}\Theta_{A}(t)\sum_{k_A}s_{k_A}^A(t)k_AP_{A}(k_A), \eqno (\text{S}4)\\$$

$$\frac{d\rho_{k_B}^{B}(t)}{dt}  =  s_{k_B}^B(t)\beta_{B}k_B\Theta_{B}(t)
-\rho_{k_B}^{B}(t), \eqno (\text{S}5)$$

$$\frac{dr_{k_B}^{B}(t)}{dt}  =  \rho_{k_B}^{B}(t), \eqno (\text{S}6) $$

$$\frac{dv_{k_B}^{B}(t)}{dt} =  p\beta_{A}\Theta_{A}(t)\sum_{k_A}s_{k_A}^A(t)
k_AP_{A}(k_A), \eqno (\text{S}7)$$
where $\Theta_{A}(t)$ [$\Theta_{B}(t)$] is the probability that a neighboring node
in layer A (layer B) is in the infected state.

From Eqs.~(S1)-(S7), the density associated with each
distinct state in layer $A$ or $B$ is given by

$$
X_{h}(t) = \sum_{k_{h}=1}^{k_{h,max}}P_{h}(k_h)X_{k_h}^{h}(t). \eqno (\text{S}8)
$$
where $h\in\{A,B\}$, $X\in\{S,I,R,V\}$, and $k_{h,max}$ denotes the largest
degree of layer $h$. The final densities of the whole system can be obtained
by taking the limit $t\rightarrow\infty$.\\

\noindent${\Large\textbf{B. Linear analysis for the information threshold}}$\\

On an uncorrelated layered network, at the outset of the spreading dynamics,
the whole system can be regarded as consisting of two coupled SI-epidemic
subsystems~\cite{Newman11:2010} with the time evolution
described by Eqs.~(S2) and (S5). For $t \rightarrow 0$, we have
$s_{k_A}^A(t) \approx 1$ and $s_{k_B}^B(t) \approx1 $, which reduce
Eqs.~(S2) and (S5) to
$$
\begin{cases}
    \frac{d\rho_{k_A}^{A}(t)}{dt} =  \beta_{A}k_A\Theta_{A}(t) +
\beta_{B}\langle k_B\rangle\Theta_{B}(t)-\rho_{k_A}^A(t), \nonumber\\
\frac{d\rho_{k_B}^{B}(t)}{dt}  = \beta_{B}k_B\Theta_{B}(t)-\rho_{k_B}^{B}(t).
\end{cases}  \eqno (\text{S}9)
$$
For convenience, Eq.~(S9) can be written concisely as
$$
\frac{d\vec{\rho}}{dt}=C\vec{\rho}-\vec{\rho}, \eqno (\text{S}10)
$$
where the vector of infected density is defined as
\begin{displaymath}
\vec{\rho} \equiv (\rho_{k_A=1}^{A},\ldots,\rho_{k_{A,max}}^{A},\rho_{k_B=1}^{B},
\ldots,\rho_{k_{B,max}}^{B})^{T},
\end{displaymath}
and $C$ is a block matrix in the following form:
$$ \label{eq:C_block}
C=\left(
\begin{array}{ccc}
    C^A & D^{B}\\
    0 & C^B\\
  \end{array}
\right), \eqno (\text{S}11)
$$
with matrix elements given by
\begin{eqnarray}
\nonumber
C_{k_A,k'_A}^A & = & [\beta_A{k_A}({k'_A}-1)P_{A}(k'_A)]/{\langle k_{A}\rangle},
\\ \nonumber
C_{k_B,k'_B}^B & = & [\beta_B{k_B}({k'_B}-1)P_{B}(k'_B)]/{\langle k_{B}\rangle},
\\ \nonumber
D_{k_B,k'_B}^B & = & \beta_B({k'_B}-1)P_{B}(k'_B).
\end{eqnarray}
In general, information spreading on layer $A$ can be facilitated by the
outbreak of the epidemic on layer $B$, as an infected node in layer $B$
instantaneously makes its counterpart node in layer $A$ ``infected'' with
the information about the disease. This coupling effect, in combination with
the intrinsic spreading dynamics on layer $A$, leads to more informed
nodes in the communication layer than infected nodes on layer $B$. If the
maximum eigenvalue $\Lambda_C$ of matrix $C$ is greater than $1$, an outbreak
of the information will occur in the system~\cite{Saumell-Mendiola11:2012}.
We then have
$$
\Lambda_C=\mbox{max}\{\Lambda_A, \Lambda_B\}, \eqno~(\text{S}12)
$$
where $\mbox{max}\{\}$ denotes the greater of the two, and
\begin{eqnarray}
\nonumber
\Lambda_A & = & \beta_A(\langle k_A^2\rangle-\langle k_A\rangle)/\langle k_A\rangle,
\\ \nonumber
\Lambda_B & = & \beta_B(\langle k_B^2\rangle-\langle k_B\rangle)/\langle k_B\rangle
\end{eqnarray}
are the maximum eigenvalues of matrices $C^A$ and $C^B$~\cite{Mieghem11:2011},
respectively. The outbreak threshold of information spreading in layer $A$
is given by
$$
\beta_{Ac} = \left\{\begin{array}{l}\beta_{Au},
\quad for \quad \beta_{B}\leq \beta_{Bu} \\
0, \quad \quad for  \quad \beta_{B}> \beta_{Bu}
\end{array} \right. \eqno~(\text{S}13)
$$
where $\beta_{Au}   \equiv   \langle k_A\rangle/({\langle k_A^2\rangle
-\langle k_A\rangle}$ and  $\beta_{Bu}   \equiv
 \langle k_B\rangle/(\langle k_B^2\rangle-\langle k_B\rangle$
denote the outbreak threshold of information spreading on layer $A$ when it
is isolated from layer $B$, and that of epidemic spreading on layer $B$ when
the coupling between the two layers is absent, respectively.\\

\noindent${\Large\textbf{C. Competing percolation theory for epidemic threshold}}$\\

To elucidate the interplay between epidemic and vaccination spreading,
we must first determine which one is the faster ``disease.'' At the early time
of the epidemic outbreak on the isolated layer $B$, the average number of
infected nodes grows exponentially as
$$N_{e}=n_0R_{e}^{t}=n_0e^{t\ln R_{e}}, \eqno~(\text{S}14)
$$
where $R_{e}=\beta_{B}/\beta_{Bu}$ is the basic reproductive number for the
disease on the isolated layer $B$~\cite{Newman11:2002-2}, and $n_0$ denotes the
number of initially infected nodes. Similarly, for information spreading on
the isolated layer $A$, the average number of informed nodes at the early
time is
$$N_{i}=n_0R_{i}^{t}=n_0e^{t\ln R_{i}}, \eqno~(\text{S}15)
$$
where $R_{i}=\beta_{A}/\beta_{Au}$ is the reproductive number for
information spreading on the isolated layer $A$. The resulting number of
vaccinated nodes on layer $B$ is
$$
N_{v}=pn_0R_{i}^{t}=pn_0e^{t\ln R_{i}}. \eqno~(\text{S}16)
$$
Since both epidemic and vaccination spreading processes exhibit exponential
growth, we can obtain the ratio of their growth rates as
$$
\theta=\frac{R_e}{R_{i}}=\frac{\beta_B\beta_{Au}}{\beta_A\beta_{Bu}}. \eqno~(\text{S}17)
$$
For $\theta>1$, i.e., $\beta_B\beta_{Au}>\beta_A\beta_{Bu}$, the epidemic
disease spreads faster than the vaccination. In this case, the vaccination spread
is insignificant and can be neglected.

To uncover the impact of information spreading on epidemic outbreak,
we focus on the case of faster vaccination, i.e., $\theta<1$, in accordance
with the fact that information always tends to spread much faster than
epidemic in a modern society. Given that vaccination and epidemic can
be treated successively and separately, the threshold of epidemic outbreak
can be derived by a bond percolation analysis~\cite{Newman11:2005,Karrer11:2011}.

Firstly, when information spreading on layer $A$ is over, the density of
informed population is given by~\cite{Newman11:2002-2}
$$
S_A=1-G_{A0}(u), \eqno~(\text{S}18)
$$
where $G_{A0}(x)=\sum_{k_A}P_A(k_A)x^{k_A}$ is the generating function for
the degree distribution of layer $A$, and $u$ is the probability that a node
is not connected to the giant cluster via a particular one of its edges, which
can be solved by
$$
u = 1 - \beta_A + \beta_AG_{A1}(u), \eqno~(\text{S}19)
$$
where $G_{A1}(x)=\sum_{k_A}Q_A(k_A)x^{k_A}$ is the generating function for
the excess degree distribution, $Q_A(k_A)=(k_A+1)P_A(k_A+1)/\langle k_A\rangle$,
of layer $A$. Since $p$ is the probability that an informed node in layer
$A$ makes its counterpart node in layer $B$ vaccinated, the number of
vaccinated or removed nodes in layer $B$ is $pS_A$. A necessary condition
for the outbreak of epidemic is the existence of a giant residual cluster in
layer $B$~\cite{Gao11:2012}. We have
$$
1 - pS_A > f_{Bc} = \frac{1}{G_{B1}'(1)}, \eqno~(\text{S}20)
$$
where $G_{B1}(x)=\sum_{k_B}Q_B(k_B)x^{k_B}$ is the generating function for the
excess degree distribution, $Q_B(k_B)=(k_B+1)P_B(k_B+1)/\langle k_B\rangle$,
of layer $B$, and the prime denotes derivative. From Eq.~(S20), we see
that epidemic outbreak can occur only if $pS_A < 1 - 1/G_{B1}'(1)$.

The degree distribution of the residual network of layer $B$ is given
by~\cite{Cohen11:2001,Pastor-Satorras11:2002}
$$
\widetilde{P}_{B}(\widetilde{k}_B) = f\sum_{k_{B}' =
\widetilde{k}_B}^{\infty}P_{B}(k_{B}')\binom{k_{B}'}{\widetilde{k}_B}
(1-f)^{k_{B}'-\widetilde{k}_B}f^{\widetilde{k}_B}, \eqno~(\text{S}21)
$$
where $f=1-pS_A$ is the probability that a node is in the residual network.
The generating function for the degree distribution of the residual network
is then~\cite{Newman11:2005}
$$
H_{B0}(x) = fG_{B0}(1-f+fx), \eqno~(\text{S}22)
$$
where $G_{B0}(x)=\sum_{k_B}P_B(k_B)x^{k_B}$ is the generating function for
the degree distribution of layer $B$. The generating function for its excess
degree distribution is
$$
H_{B1}(x) = \frac{H_{B0}^{\prime}(x)}{H_{B0}^{\prime}(1)}. \eqno~(\text{S}23)
$$
The basic reproductive number for a disease spreading over the residual network
of layer $B$ is then given by~\cite{Newman11:2002-2}
$$
\widetilde{R}_{i} = \beta_B H_{B1}^{\prime}(1). \eqno~(\text{S}24)
$$
The epidemic threshold corresponds to the point $\widetilde{R}_{i}=1$, and
thus we have $\beta_{Bc}=1/H_{B1}^{\prime}(1)$. From Eqs.~(S22)-(S24),
we obtain the epidemic threshold $\beta_{Bc}$ as

$$
\beta_{Bc} = \frac{\langle k_{B}\rangle}{(1-pS_A)
(\langle k_{B}^{2}\rangle-\langle k_{B}\rangle)}, \eqno~(\text{S}25)
$$
where $S_A$ is the density of the informed population, which
can be obtained by solving Eqs.~(S18)~and~(S19).\\
\newpage
\noindent${\Large\textbf{S2. Spreading dynamics on correlated double-layer networks}}$\\

We assume that layer $A$ has the same degree distribution as layer $B$.
After a certain fraction $q$ of pairs of nodes, one from each layer,
have been randomly rematched, the conditional probability $P(k_B|k_A)$
can be written as
$$P(k_B|k_A) = qP_B(k_B) + (1-q)\delta_{k_B,k_A}, \eqno (\text{S}26) $$

or
$$
P(k_A|k_B)=qP_A(k_A)+(1-q)\delta_{k_A,k_B}. \eqno (\text{S}27)
$$

\noindent${\Large\textbf{A. Mean-field rate equations}}$\\

Using Eqs.~(S1)-(S3), we can write the mean-field rate equations
for information spreading on layer $A$ as

$$\frac{ds_{k_A}^A(t)}{dt} =-s_{k_A}^A(t)\{\beta_{A}k_A\Theta_{A}(t)
+\beta_{B}\Theta_{B}(t)\sum_{k_B}k_B [qP_{B}(k_B)+(1-q)\delta_{k_B,k_A}]\}, \eqno (\text{S}28)$$

$$
\frac{d\rho_{k_A}^{A}(t)}{dt}=s_{k_A}^A(t)\{\beta_{A}k_A\Theta_{A}(t)
+\beta_{B}\Theta_{B}(t)\sum_{k_B}k_B [qP_{B}(k_B)+(1-q)\delta_{k_B,k_A}]\}-\rho_{k_A}^{A}(t), \eqno (\text{S}29)
$$

$$\frac{dr_{k_A}^A(t)}{dt} =  \rho_{k_A}^A(t). \eqno (\text{S}30)$$

Similarly, the mean-field rate equations for epidemic spreading on layer
$B$ are

$$
\frac{ds_{k_B}^B(t)}{dt}=-s_{k_B}^B(t)\beta_{B}k_B\Theta_{B}(t)
-p\beta_{A}\Theta_{A}(t) \\
\sum_{k_A}s_{k_A}^A(t)k_A[qP_A(k_A)+(1-q)\delta_{k_A,k_B}], \eqno (\text{S}31)
$$

$$
\frac{d\rho_{k_B}^{B}(t)}{dt} =  s_{k_B}^B(t)\beta_{B}k_B\Theta_{B}(t)
-\rho_{k_B}^{B}(t), \eqno (\text{S}32)
$$

$$\frac{dr_{k_B}^{B}(t)}{dt}  =  \rho_{k_B}^{B}(t), \eqno (\text{S}33)$$

$$
\frac{dv_{k_B}^{B}(t)}{dt}  =  p\beta_{A}\Theta_{A}(t)
\sum_{k_A}s_{k_A}^A(t)k_A[qP_A(k_A)
 +  (1-q)\delta_{k_A,k_B}]. \eqno (\text{S}34)
$$
Substituting Eqs.~(S28)-(S34) into Eq.~(S8), we can get  the density associated with each
distinct state in layer $A$ or $B$.\\

\noindent${\Large\textbf{B. Linear analysis for the information threshold}}$\\

At the outset of the spreading dynamics, the whole system can be regarded as
two coupled SI-epidemic subsystems~\cite{Newman11:2010} with the time evolution
described by Eqs.~(S29) and (S32). In the limit $t \rightarrow 0$,
 we have $s_{k_A}^A(t) \approx 1$ and
$s_{k_B}^B(t) \approx 1$. Equations~(S29) and (S32) can then
be reduced to
$$
\begin{cases}
    \frac{d\rho_{k_A}^{A}(t)}{dt}  = \beta_{A}k_A\Theta_{A}(t)
    +\beta_{B}[q\langle k_B\rangle+(1-q)k_A]\Theta_{B}(t)
    -\rho_{k_A}^A(t),\\
\frac{d\rho_{k_B}^{B}(t)}{dt}=\beta_{B}k_B\Theta_{B}(t)-\rho_{k_B}^{B}(t).
\end{cases}  \eqno (\text{S}35)
$$
which can be written concisely as
$$
\frac{d\vec{\rho}}{dt}=C\vec{\rho}-\vec{\rho}, \eqno (\text{S}36)
$$
where the matrix $C$ has the same form as in Eq.~(S11) and
\begin{eqnarray}
\nonumber
C_{k_A,k'_A}^A & = & [\beta_A{k_A}({k'_A}-1)P_{A}(k'_A)]/{\langle k_{A}\rangle},
\\ \nonumber
C_{k_B,k'_B}^B & = & [\beta_B{k_B}({k'_B}-1)P_{B}(k'_B)]/{\langle k_{B}\rangle},
\\ \nonumber
D_{k_B,k'_B}^B & = & \beta_B[q\langle k_B\rangle
+(1-q)k_A]({k'_B}-1)P_{B}(k'_B)/\langle k_B\rangle.
\end{eqnarray}
The threshold of information outbreak is given by
$$
\beta_{Ac} = \left\{\begin{array}{l}\beta_{Au},
\quad for \quad \beta_{B}\leq \beta_{Bu}, \\
0, \quad \quad for  \quad \beta_{B}> \beta_{Bu},
\end{array} \right. \eqno (\text{S}37)
$$
which is the same as Eq.~(9) in the main text. As described in
uncorrelated networks, there are two distinct mechanisms that can
lead to the outbreak of information on layer $A$, and these hold for
correlated layered-networks as well. For $\beta_{B}\leq \beta_{Bu}$, only a
small number of nodes in layer $B$ are infected, so the impact of the
disease on information-outbreak threshold on layer $A$ is negligible.
For $\beta_{B}> \beta_{Bu}$, epidemic spreading can result in the outbreak
of information. In this case, the information-outbreak threshold is zero.\\

\noindent${\Large\textbf{C. Competing percolation theory for epidemic threshold}}$\\

For $\beta_{A}\leq\beta_{Au}$, information itself cannot spread through
the population. There is thus hardly any effect of the information layer
on the epidemic spreading on layer $B$, and we have $\beta_{Bc}\approx\beta_{Bu}$.
But for $\beta_{A}>\beta_{Au}$, the effect of information spreading on
the epidemic threshold cannot be ignored. To assess quantitatively the
influence, we focus on the case of faster information spread, i.e.,
$\beta_A\beta_{Bu}>\beta_B\beta_{Au}$, rendering applicable a bond
percolation analysis similar to uncorrelated networks.
Specifically, after information spreads on layer $A$, the percentage
of nodes that get the information is $S_A$, and the density of recovered
nodes of degree $k_A$ is $r_{k_A}^A=1-u^{k_A}$, where $u$ is the probability
that a node is not connected to the giant cluster by a particular
edge [Eq.~(S19)]. Vaccinating a number of counterpart nodes
results in the random removal of some edges which connect the vaccinated
nodes with the remaining nodes~\cite{Cohen11:2001,Pastor-Satorras11:2002}.
The probability $\widetilde{h}$ of an edge linking to a vaccinated node is
$$
\widetilde{h} = \frac{p\sum_{k_B}[(1-q)r_{k_A}+qS_A]k_BP_{B}(k_B)}
{\langle k_{B}\rangle}. \eqno (\text{S}38)
$$
The new degree distribution of the residual network on layer $B$ is
thus given by
$$
\widetilde{P}_{B}(\widetilde{k}_B)  =
\sum_{k_{B}'=\widetilde{k}_B}^{\infty}\{1-p[(1-q)r_{k_A}+qS_A]\}
 P_{B}(k_{B}')\binom{k_{B}'}{\widetilde{k}_B}
(1-\widetilde{h})^{\widetilde{k}_B}{\widetilde{h}}^{k_{B}'-\widetilde{k}_B}. \eqno (\text{S}39)
$$
The requirement that a giant residual cluster exists is
$$
\frac{\langle{\widetilde{k}_{B}}^{2}\rangle}{\langle \widetilde{k}_{B}\rangle}>2, \eqno (\text{S}40)
$$
where $\langle \widetilde{k}_{B}\rangle$ and $\langle{\widetilde{k}_{B}}^{2}\rangle$
are the first and second moments of the degree distribution, respectively.
Finally, we obtain the epidemic threshold as
$$ \label{beta_BCCor}
\beta_{Bc}=\frac{\langle \widetilde{k}_{B}\rangle}
{\langle \widetilde{k}_{B}^{2}\rangle-\langle \widetilde{k}_{B}\rangle}. \eqno (\text{S}41)
$$

\newpage
\noindent${\Large\textbf{S3. Simulation results
}}$\\

We first describe the simulation process of the two spreading dynamics on double-layer networks,
and then demonstrate the validity of the theoretical analysis on uncorrelated networks with
different network sizes and degree exponents, finally, we present results for SF-SF correlated networks.\\

\noindent${\Large\textbf{A. Simulation process}}$\\

To initiate an epidemic spreading process, a node in layer $B$ is randomly
infected and its counterpart node in layer $A$ is thus in the informed state,
too. The updating process is performed with parallel dynamics, which is widely
used in statistical physics~\cite{Marro11:1999}. At each time step, we first
calculate the informed (infected) probability $\pi_{A}=1-(1-\beta_A)^{n_I^A}$ [$\pi_B=1-(1-\beta_B)^{n_I^B}$]
that each susceptible node in layer $A$ ($B$) may be informed (infected) by
its informed (infected) neighbors, where $n_I^A$ ($n_I^B$) is the number of its
informed (infected) neighboring nodes.

According to the dynamic mechanism, once node $A_i$ is in the susceptible state,
its counterpart node $B_i$ will be also in the susceptible state.
Considering the asymmetric coupling between the two layers in this case,
both the information-transmission and disease-transmission events
can hardly occur at the same time. Thus, with probability $\pi_A/(\pi_A+\pi_B)$,
node $A_i$ have a probability $\pi_A$ to get the information from its informed neighbors in layer $A$.
If node $A_i$ is informed, its counterpart node $B_i$
will turn into the vaccination state with probability $p$.
With probability $\pi_B/(\pi_A+\pi_B)$, node $B_i$ have a probability $\pi_B$
to get the infection from its infected neighbors in layer $B$,
and then node $A_i$ also get the information about
the disease.

In the other case that node $B_i$ and its corresponding node $A_i$
are in the susceptible state and the informed (or refractory) state respectively,
only the disease-transmission event can occur at the time step.
Thus, node $B_i$ will be infected with probability $\pi_B$.

After renewing the states of susceptible nodes,
each informed (infected) node can enter the recovering phase with probability
$\mu_A=1.0$ ($\mu_B=1.0$). The spreading dynamics terminates when all informed (or infected) nodes in both
layers are recovered, and the final densities $R_A$, $R_B$, and $V_B$ are
then recorded. The simulations are implemented using $30$ different two-layer network realizations.
The network size of $N_A=N_B=2 \times 10^4$ and average degrees
$\langle k_A\rangle=\langle k_B\rangle=8$ are used for all
subsequent numerical results, unless otherwise specified.\\

\noindent${\Large\textbf{B. Uncorrelated double-layer networks}}$\\

\begin{figure}
\begin{center}
\renewcommand{\figurename}{FIG. S}
\epsfig{file=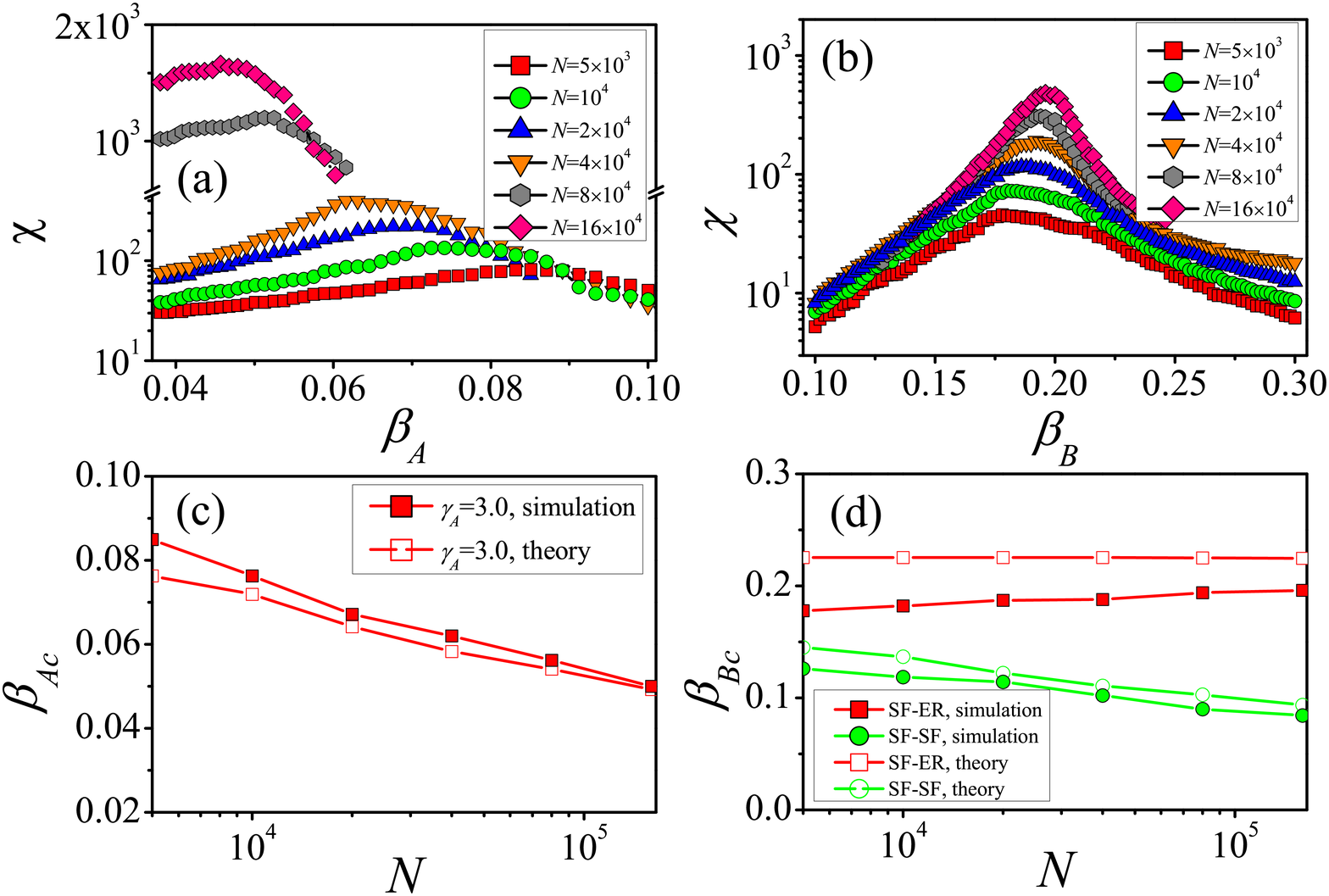,width=1\linewidth}
\caption{(Color online) On SF-ER networks, the susceptibility measure
$\chi$  as a function of the information-transmission rate $\beta_A$ at
$\beta_B=0.1$ (a) and the disease-transmission rate $\beta_B$ at $\beta_A=0.3$ (b)
for $N=5\times10^3$ (red squares), $N=10^4$ (green circles),
$N=2\times10^4$ (blue up triangles), $N=4\times10^4$ (orange down triangles),
$N=8\times10^4$ (gray hexagons) and $N=16\times10^4$ (pink diamonds);
(c) the information threshold $\beta_{Ac}$ as a function of network
size $N$ at $\beta_B=0.1$. (d) The epidemic threshold $\beta_{Bc}$
as a function of $N$ at $\beta_A=0.3$ for
SF-ER networks (red solid squares) and SF-SF networks
(green solid circles). The same hollow symbols represent the corresponding
theoretical thresholds. The other parameters are the degree exponent
$\gamma_A=3.0$ (or $\gamma_B=3.0$) and vaccination rate $p=0.5$.}
\label{figS1}
\end{center}
\end{figure}

\begin{figure}
\begin{center}
\renewcommand{\figurename}{FIG. S}
\epsfig{file=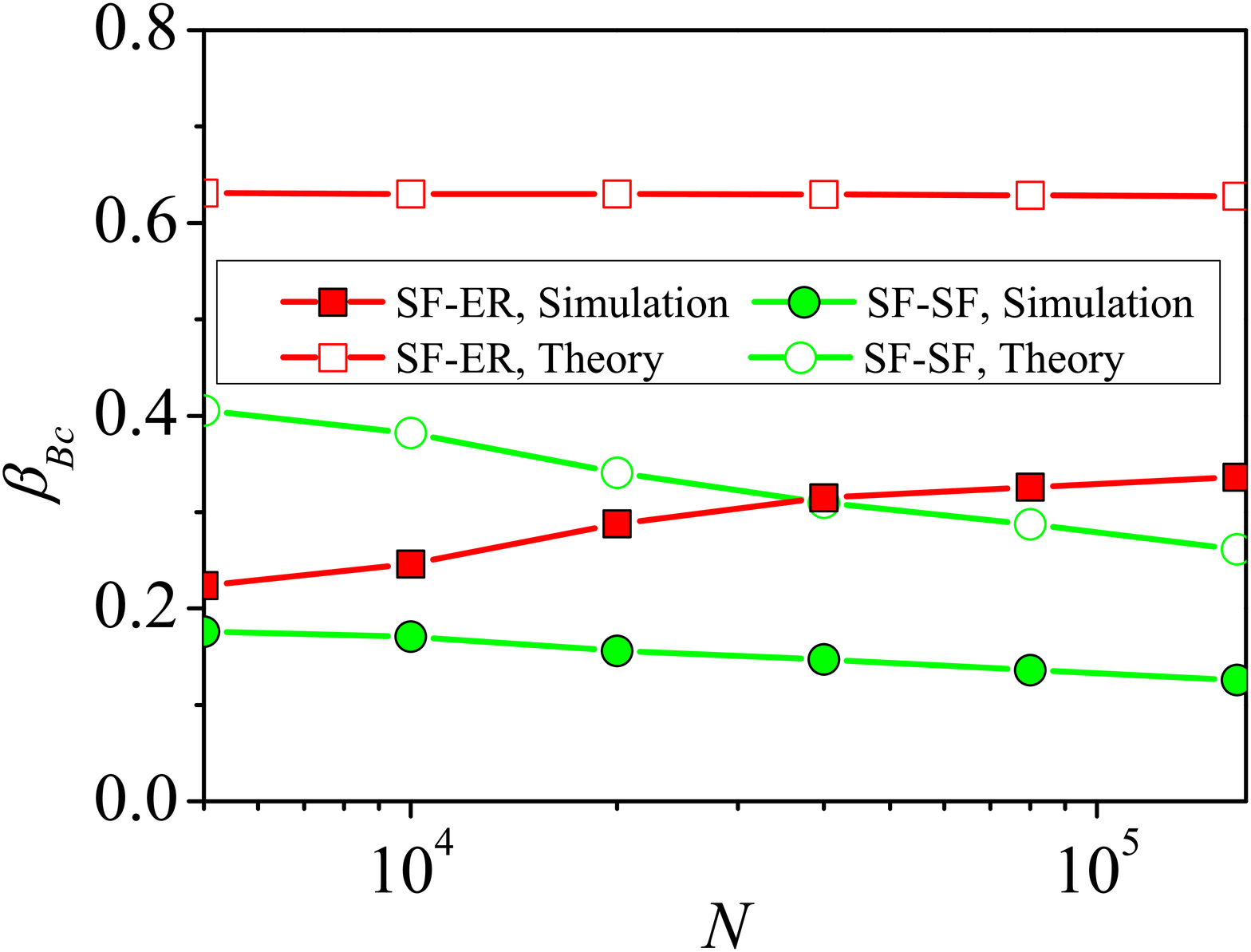,width=0.7\linewidth}
\caption{(Color online) The epidemic threshold $\beta_{Bc}$
as a function of network size $N$ for SF-ER networks (red solid squares) and SF-SF networks
(green solid circles) at $\beta_A=0.3$ and $p=0.9$. The same hollow symbols represent the corresponding
theoretical thresholds. For each SF layer, the degree exponent is set to
$\gamma_A=3.0$ (or $\gamma_B=3.0$).}
\label{figS2}
\end{center}
\end{figure}

The effect of network size $N$ on the information and
epidemic outbreak thresholds is first studied. According to Eq.~(S13),
the behavior of the information threshold can be classified into two classes.
For $\beta_B\leq\beta_{Bu}$, the disease transmission on layer $B$
has little impact on the information threshold, as we have
$\beta_{Ac}\approx\beta_{Au}= \langle k_A\rangle/({\langle k_A^2\rangle -
\langle k_A\rangle})$; while $\beta_{Ac}=0.0$ for $\beta_B>\beta_{Bu}$.
We here focus on the information threshold for
$\beta_B\leq\beta_{Bu}$. From Figs. ~S\ref{figS1}(a) and (c),
we see that the theoretical predictions are basically accordant with
the simulated thresholds for different network sizes. With the growth of
network size, the information threshold decreases as $\langle k^2\rangle$ of
layer $A$ increases~\cite{Boguna11:2013}. According to Eq.~(S25),
the theoretical epidemic threshold can be predicted.
For SF-ER double-layer networks, Figs.~S\ref{figS1}(b) and (d)
shows that the simulated epidemic thresholds deviate slightly from the theoretical predictions.
However, the larger deviations occur for the larger values of the vaccination rate $p$,
e.g., $p=0.9$ in Fig.~S\ref{figS2},
because the basic assumption of competing percolation theory
is not strictly correct for the finite-size networks.
As pointed out by Karrer and Newman~\cite{Karrer11:2011}, in the limit of large network size $N$,
the vaccination and epidemic processes can be treated successively and separately.
On the double-layer networks with finite network size,
the effect of information spreading is somewhat over-emphasized.
From Figs.~S\ref{figS1} and S\ref{figS2},
we also see that the discrepancy between the simulated and theoretical thresholds
decreases with network size $N$.

\begin{figure}
\begin{center}
\renewcommand{\figurename}{FIG. S}
\epsfig{file=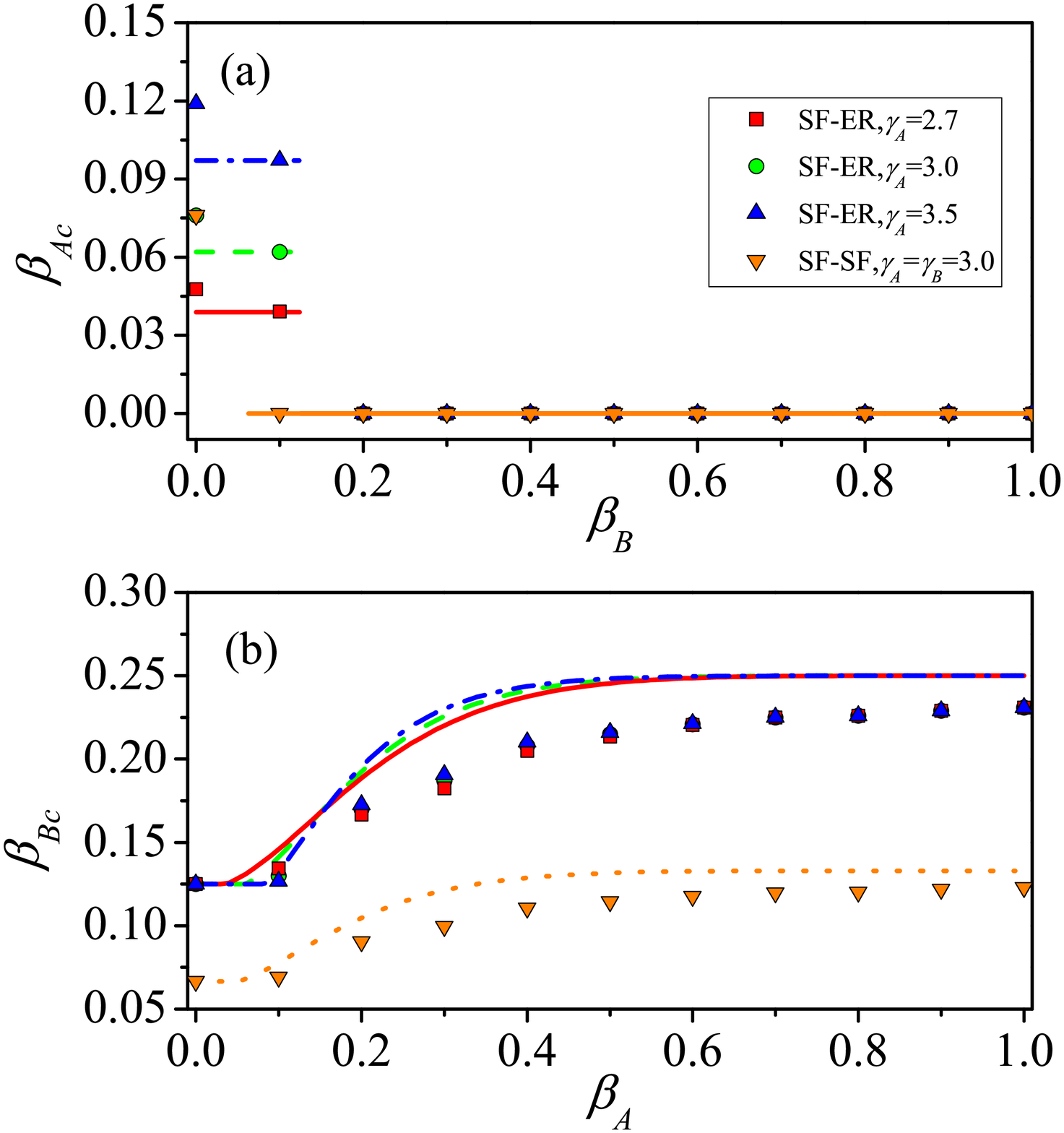,width=0.7\linewidth}
\caption{(Color online) On various double-layer networks,
$\beta_{Ac}$ versus $\beta_B$ (a) and $\beta_{Bc}$ versus $\beta_A$ (b)
for the SF-ER networks with $\gamma_A=2.7$ (red squares), the
SF-ER networks with $\gamma_A=3.0$ (green circles), the SF-ER networks with
$\gamma_A=3.5$ (blue up triangles) and the SF-SF networks with $\gamma_A=\gamma_B=3.0$
(orange down triangles). The analytical predictions of $\beta_{Ac}$ and $\beta_{Bc}$
are from Eq.~(S13) and Eq.~(S25), respectively. The vaccination rate is set to $p=0.5$.
}
\label{figS3}
\end{center}
\end{figure}

We then investigate how the degree heterogeneity of layer $A$ influences the information and
epidemic outbreak thresholds by adjusting the exponent $\gamma_A$.
The information thresholds for the different exponents of layer $A$
are compared in Fig.~S\ref{figS3}(a), and the stronger heterogeneity of layer $A$
(i.e., smaller $\gamma_A$) can more easily make the information outbreak.
Fig.~S\ref{figS3}(b) shows that increasing the heterogeneity of layer $A$
can slightly raise the epidemic threshold $\beta_{Bc}$ at a small information-transmission rate $\beta_A$,
while making for the epidemic outbreak at a large $\beta_A$.
This phenomenon results from the different effects of the heterogeneity
on the information spreading under different transmission rates.
The more homogeneous degree distribution does not always
hinder the diffusion of information,
especially at a large transmission rate~\cite{Pastor-Satorras11:2001,Pastor-Satorras11:2002}.

\begin{figure}
\begin{center}
\renewcommand{\figurename}{FIG. S}
\epsfig{file=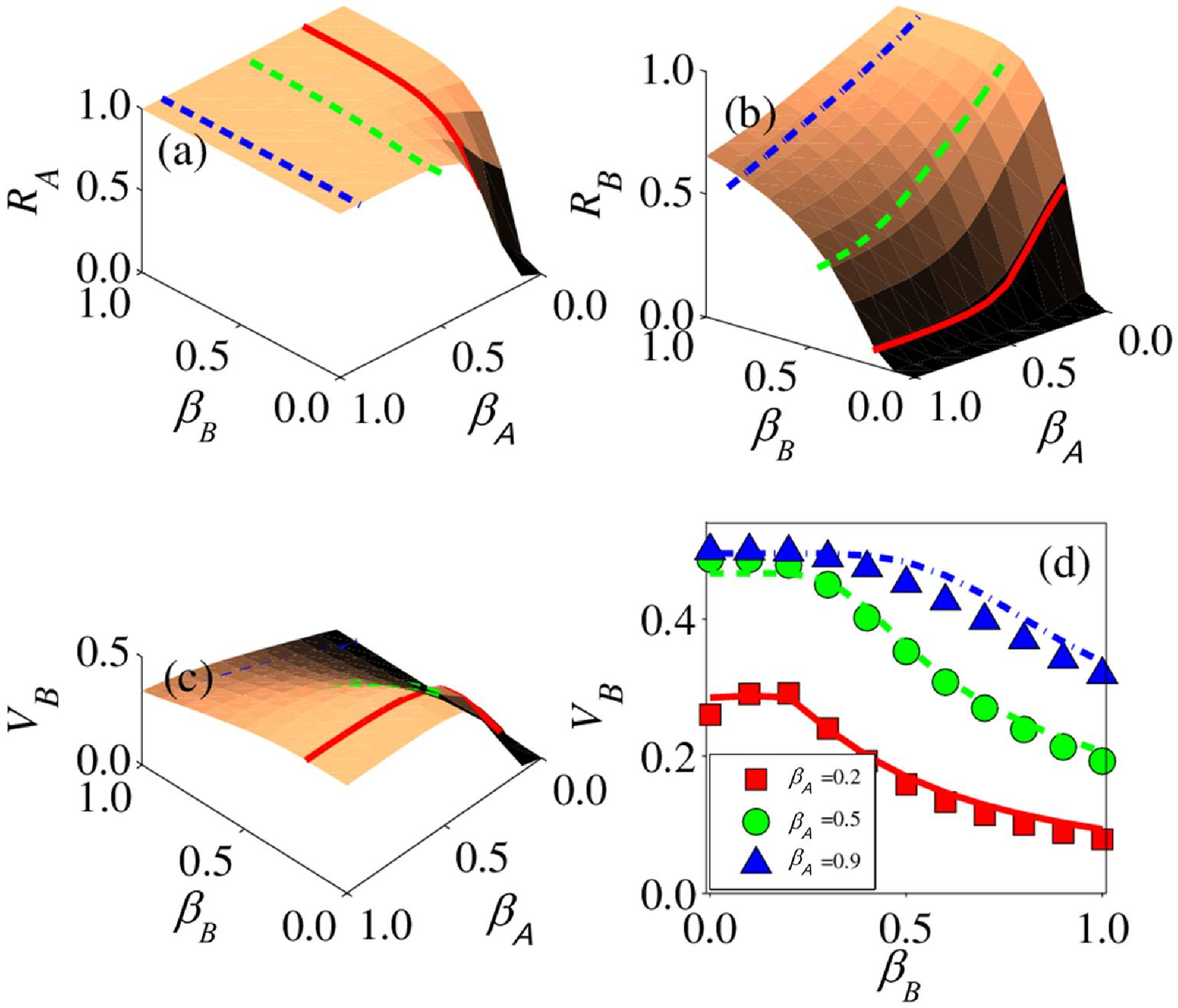,width=0.7\linewidth}
\caption{(Color online) For SF-SF networks, the final density in each state
versus the parameters $\beta_A$ and $\beta_B$: (a) recovered density $R_A$,
(b) recovered density $R_B$, (c) the vaccination density $V_B$, and (d)
$V_B$ versus $\beta_B$ for $\beta_A=0.2,0.5,0.9$. The other parameter are $p=0.5$
and $\gamma_A=\gamma_B=3.0$. Different lines are the numerical solutions of
Eqs.~(S1)-(S8) in the limit $t\rightarrow\infty$.
In (a) and (d), we select three different values of $\beta_A$ (0.2, 0.5, and
0.9), corresponding to the red solid, green dashed, and blue dot-dashed lines,
respectively. In (b) and (c), three different values of $\beta_B$ are chosen
(0.2, 0.5, and 0.9), corresponding to the red solid, green dashed, and blue
dot-dashed lines, respectively. }
\label{figS4}
\end{center}
\end{figure}

To further demonstrate the validity of the theoretical analysis,
we consider the case of SF-SF double-layer networks.
Similar to the case of SF-ER networks, the gap between
the theoretical and simulated thresholds is narrowing
with the increase of network size [see Figs.~S\ref{figS1}(d) and S\ref{figS2}],
which implies the reasonability of the assumption in the thermodynamic limit.
The final dynamical state of the SF-SF spreading system is also shown
in Fig.~S\ref{figS4}, and it displays a similar phenomenon to the case of SF-ER networks.
We also see that the theoretical predictions from mean-field rate
equations are in good agreement with the simulation results.\\

\noindent${\Large\textbf{C. Correlated double-layer networks
}}$\\

\begin{figure}
\begin{center}
\renewcommand{\figurename}{FIG. S}
\epsfig{file=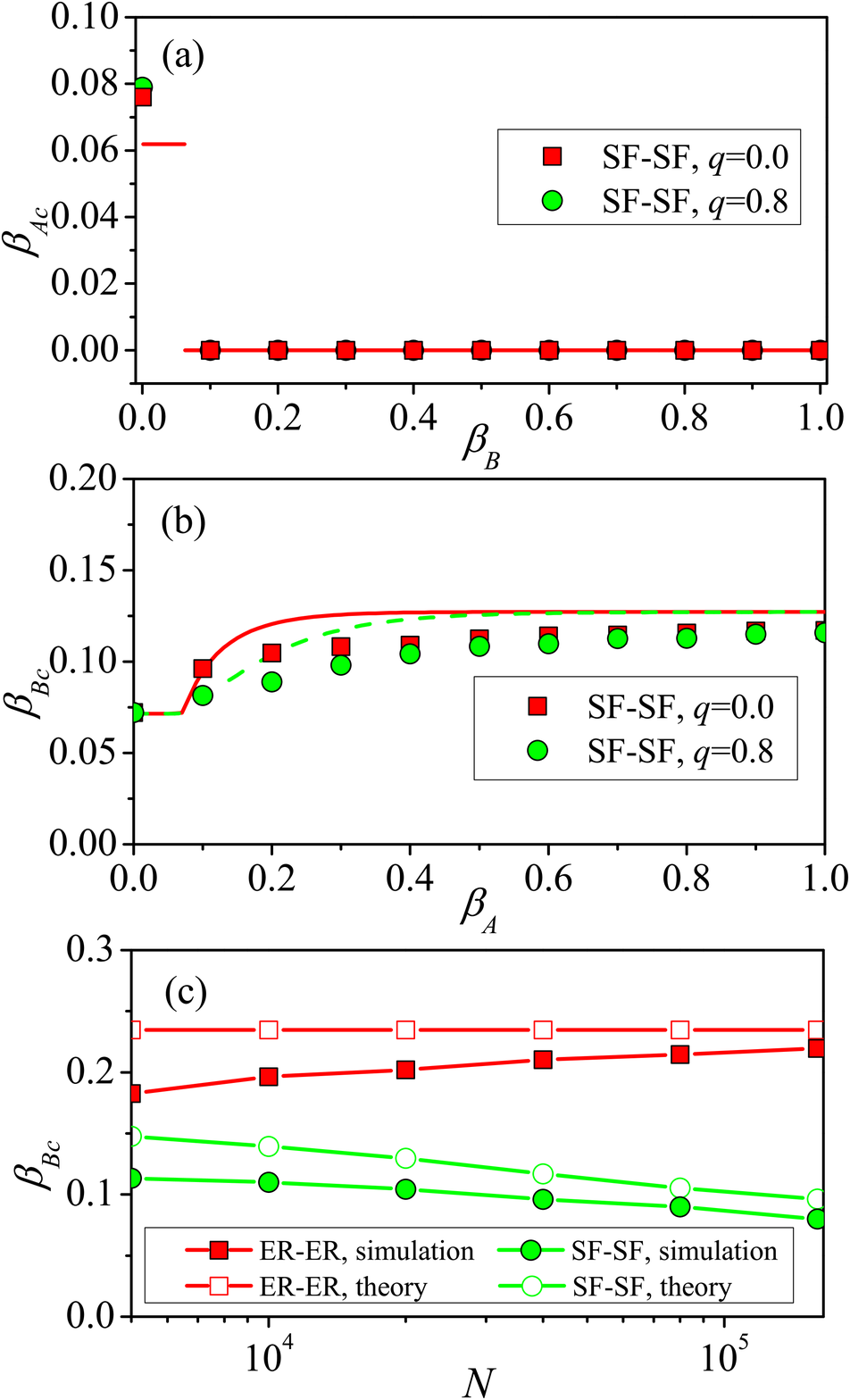,width=0.7\linewidth}
\caption{(Color online) On double-layer networks, (a)
$\beta_{Ac}$ versus $\beta_B$, (b) $\beta_{Bc}$ versus $\beta_A$,
and (c) $\beta_{Bc}$ versus $N$ at $\beta_{A}=0.3$.
In (a) and (b), red solid squares and green solid circles respectively
denote the simulation results for $q=0.0$ and $q=0.8$ on SF-SF networks,
and the lines are the corresponding theoretical thresholds.
In (c), the value of parameter $q$ is $0.0$,
solid red squares and solid green circles respectively represent the results for
ER-ER and SF-SF networks, and the same shapes are the corresponding theoretical predictions.
The other parameter are $p=0.5$ and $\gamma_A=\gamma_B=3.0$.
}
\label{figS5}
\end{center}
\end{figure}

\begin{figure}
\begin{center}
\renewcommand{\figurename}{FIG. S}
\epsfig{file=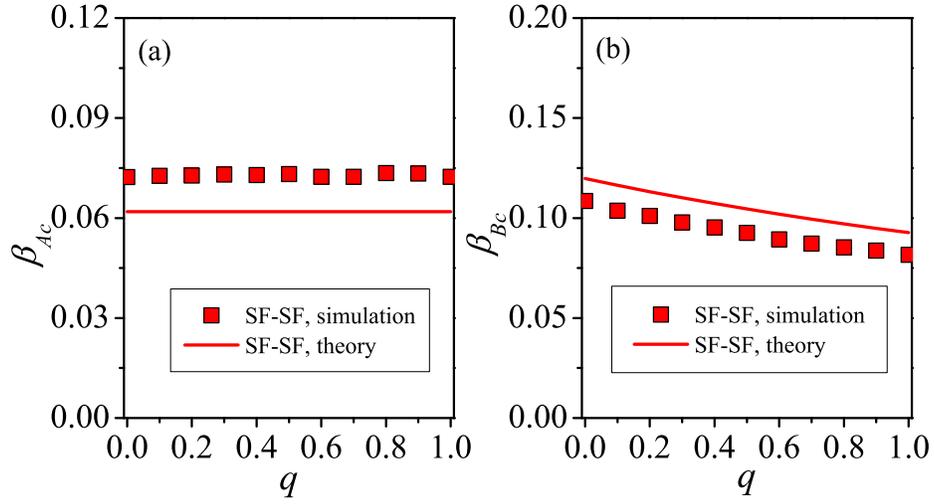,width=0.7\linewidth}
\caption{(Color online) On SF-SF networks, the effect of varying the rematching probability on outbreak
thresholds of the two types of spreading dynamics. (a) $\beta_{Ac}$ versus $q$
for $\beta_{B}=0.05$ and $p=0.5$. Red Solid line is the analytical
prediction from Eq.~(S37). (b) $\beta_{Bc}$ versus $q$
for $\beta_{A}=0.2$ and $p=0.5$. Red solid line is the analytical
prediction from Eqs.~(S38)-(S41). The value of degree exponent is $\gamma_A=\gamma_B=3.0$.}
\label{figS6}
\end{center}
\end{figure}

\begin{figure}
\begin{center}
\renewcommand{\figurename}{FIG. S}
\epsfig{file=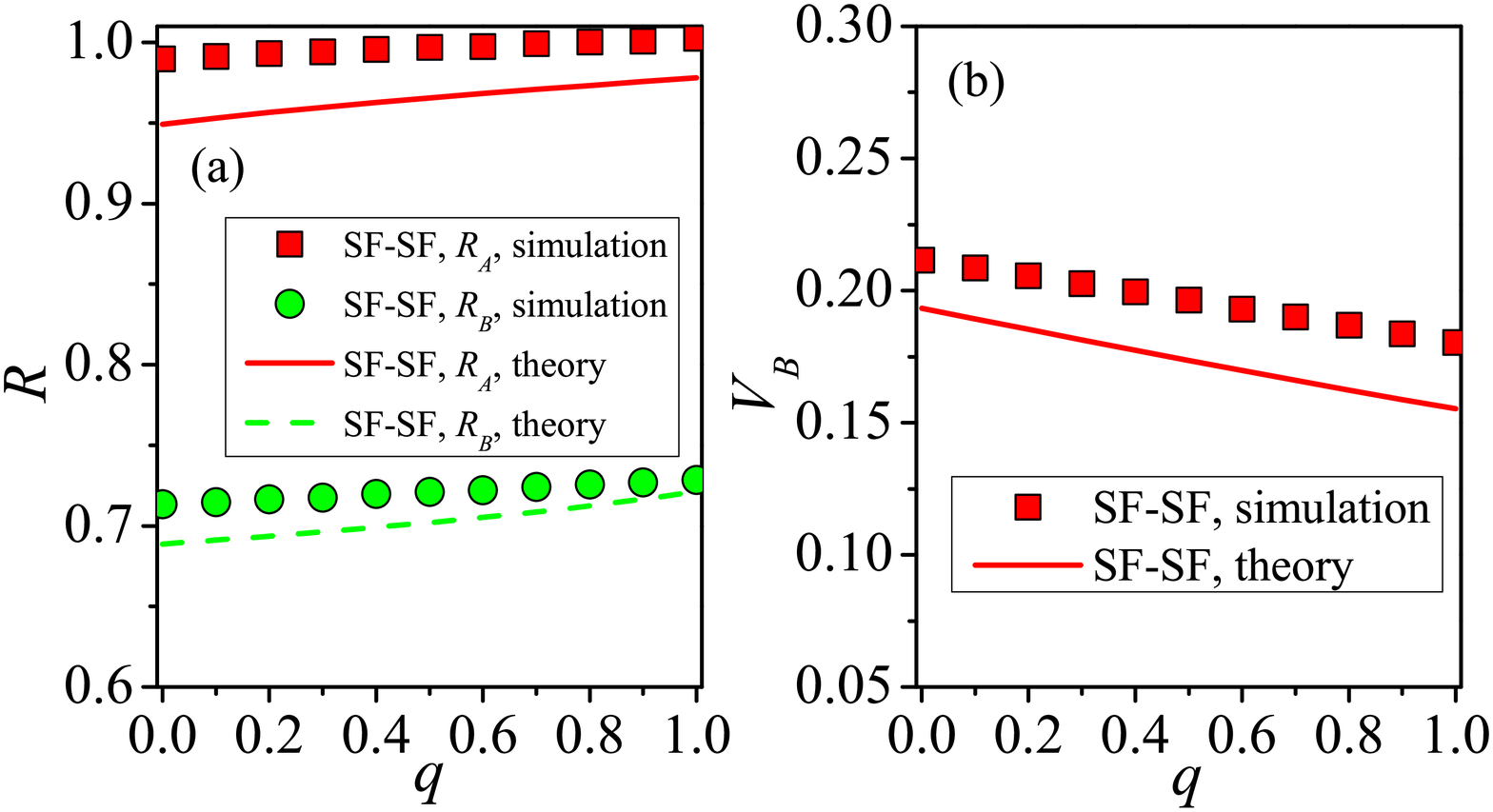,width=0.7\linewidth}
\caption{(Color online) On SF-SF networks, the effect of rematching probability on the final state.
(a) $R_{A}$ versus $q$ (red squares) and $R_{B}$ versus $q$ (green circles),
(b) $V_{B}$ versus $q$ (red squares).
Different lines represent the analytic solutions for SF-SF networks, calculated by
summing the final densities of all degrees from Eqs.~(S28)-(S34).
The parameter setting is $\gamma_A=\gamma_B=3.0, \beta_{A}=0.2$, $\beta_{B}=0.4$ and $p=0.5$.}
\label{figS7}
\end{center}
\end{figure}

On SF-SF correlated networks, we investigate the effect of positive inter-layer correlation on the two types of
spreading dynamics.
As shown in Figs. S\ref{figS5}, S\ref{figS6} and S\ref{figS7},
with the increase of the correlation $m_s$ (by reducing the rematching probability $q$),
the information threshold remains unchanged but the epidemic threshold can be enhanced,
making the contact layer more robust to epidemic outbreak,
which is consistent with the results for ER-ER correlated networks.

\end{document}